\pdfoutput=1
\documentclass[
aps,
floatfix,
letterpaper,
prx,
singlecolumn,
reprint,
superscriptaddress,
longbibliography
]{revtex4-2}

\usepackage[sectionbib]{bibunits}
\defaultbibliographystyle{naturemag}
\defaultbibliography{ENDOR_fixed} 
\makeatletter
\AtBeginDocument{\let\rtx@bibliography@first\bibliography
  \def\bibliography#1{\rtx@bibliography@first{#1}\global\let\bibliography\@gobble}}
\makeatother

\usepackage[colorlinks=true, linkcolor=blue]{hyperref}%

\usepackage{lipsum}
\usepackage{graphicx} 
\usepackage{amsfonts}
\usepackage{amsmath}
\usepackage{braket}
\usepackage{dsfont}
\usepackage[capitalize]{cleveref}
\usepackage[utf8]{inputenc}
\usepackage{comment}
\usepackage{capt-of}
\usepackage{booktabs}
\usepackage{lineno}
\newif\ifhighlight
\highlightfalse
\ifhighlight
\newcommand{\red}[1]{\textcolor{red}{#1}}

\newcommand{\blue}[1]{\textcolor{blue}{#1}}%
\newcommand{\recent}[1]{\textcolor{blue}{#1}}%
\else
\newcommand{\red}[1]{#1}

\newcommand{\blue}[1]{#1}
\newcommand{\recent}[1]{#1}
\fi

\AtBeginDocument{\definecolor{BLUE}{rgb}{0,0,1}}%
\newsavebox{\dftbox}
\crefname{suppnote}{Supplementary Note}{Supplementary Notes}
\Crefname{suppnote}{Supplementary Note}{Supplementary Notes}
\crefname{suppfigure}{Supplementary Fig.}{Supplementary Figs.}
\Crefname{suppfigure}{Supplementary Fig.}{Supplementary Figs.}
\crefname{supptable}{Supplementary Table}{Supplementary Tables}
\Crefname{supptable}{Supplementary Table}{Supplementary Tables}
\crefname{suppequation}{Supplementary Equation}{Supplementary Equations}
\Crefname{suppequation}{Supplementary Equation}{Supplementary Equations}
\crefname{equation}{Equation}{Equations}
\Crefname{equation}{Equation}{Equations}
\creflabelformat{suppequation}{(#2#1#3)}

\begin{document}
\setcounter{secnumdepth}{3}
\begin{bibunit}[naturemag]
\begin{abstract}
\red{Spin defects in diamond serve as powerful building blocks for quantum technologies, especially for quantum sensing and networking. Electron-nuclear defects formed near optically active spins such as the nitrogen-vacancy center provide a resource for multi-qubit quantum registers. However, many remain uncharacterized, limiting their control and integration in quantum devices. Here, we identify the hyperfine components\blue{, including their relative sign,} and nuclear spin species of two unknown \blue{dark} spin defects using zero-field double electron-electron resonance spectroscopy and a protocol based on electron-nuclear double resonance. \blue{New \textit{ab initio} calculations assign one defect to a known nitrogen interstitial, while the hydrogen-related defect matches no calculated or previously reported structure.} We further use this protocol to demonstrate initialization, unitary control, and long-lived coherence of the hydrogen nuclear spin qubit, with a coherence time of $1.0(3)$\,ms. These characterization and control tools establish a framework for harnessing unknown electron-nuclear defects for quantum registers.}
\end{abstract}

\title{\blue{Hyperfine spectroscopy and nuclear spin control of dark defects in diamond}}

\author{Alexander Ungar} 
\thanks{Corresponding author. Email: ungar@mit.edu}
\affiliation{Massachusetts Institute of Technology, Research Laboratory of Electronics, Cambridge, MA 02139, USA}
\affiliation{Massachusetts Institute of Technology, Department of Electrical Engineering and Computer Science, Cambridge, MA 02139, USA}
\author{Hao Tang} 
\affiliation{Massachusetts Institute of Technology, Department of Materials Science and Engineering, Cambridge, MA 02139, USA}
\author{Andrew Stasiuk} 
\affiliation{Massachusetts Institute of Technology, Research Laboratory of Electronics, Cambridge, MA 02139, USA}
\affiliation{Massachusetts Institute of Technology, Department of Nuclear Science and Engineering, Cambridge, MA 02139, USA}
\author{Bo Xing} 
\affiliation{Massachusetts Institute of Technology, Research Laboratory of Electronics, Cambridge, MA 02139, USA}
\affiliation{Quantum Innovation Centre, Agency for Science, Technology and Research, Singapore 138634, Singapore}
\author{Boning Li} 
\affiliation{Massachusetts Institute of Technology, Research Laboratory of Electronics, Cambridge, MA 02139, USA}
\affiliation{Massachusetts Institute of Technology, Department of Physics, Cambridge, MA 02139, USA}
\author{Ju Li}
\affiliation{Massachusetts Institute of Technology, Department of Materials Science and Engineering, Cambridge, MA 02139, USA}
\affiliation{Massachusetts Institute of Technology, Department of Nuclear Science and Engineering, Cambridge, MA 02139, USA}
\author{Alexandre Cooper}
\affiliation{University of Waterloo, Institute for Quantum Computing, Waterloo, ON N2L 3G1, Canada}
\author{Paola Cappellaro}
\thanks{Corresponding author. Email: pcappell@mit.edu}
\affiliation{Massachusetts Institute of Technology, Research Laboratory of Electronics, Cambridge, MA 02139, USA}
\affiliation{Massachusetts Institute of Technology, Department of Physics, Cambridge, MA 02139, USA}
\affiliation{Massachusetts Institute of Technology, Department of Nuclear Science and Engineering, Cambridge, MA 02139, USA}

\maketitle

\section*{Introduction}
Nitrogen-vacancy (NV) centers provide a promising platform for nanoscale magnetic field sensing, serving as optically addressable spins with long coherence times and efficient optical polarization at room temperature~\cite{Balasubramanian2008, Maze2008, Taylor2008, Degen2017}. As a quantum sensor, NV centers can be positioned with nanometer precision~\cite{Maletinsky2012} and exhibit single electron and nuclear spin sensitivity~\cite{Grinolds2014, Lovchinsky2016}. These are key capabilities for probing condensed matter systems~\cite{Casola2018}, and for nuclear magnetic resonance imaging of complex spin systems~\cite{Abobeih2019} down to the single molecule level~\cite{Janitz2022,Du2024}. A critical challenge for quantum-enhanced sensing with this platform is achieving individual control and readout of many interacting NV spins, for example to surpass the standard quantum limit in order to increase sensitivity~\cite{Bollinger1996, Cooper2019, Zhou2025}. Alternatively, optically inactive electron-nuclear defects\blue{, referred to herein as dark defects, which are} strongly coupled to nearby NV centers can be harnessed as auxiliary qubits, enabling larger quantum registers within the NV diffraction limit~\cite{Knowles2016, Rosenfeld2018, Degen2021}. These electron-nuclear spin defects, such as the P1 center (nitrogen substitutional defect), host an electronic spin that can be distinguished via its dipolar coupling to the NV, enabling selective and coherent control. The electronic spins can be entangled with high fidelity~\cite{Joas2025} and serve as reporter spins for amplification~\cite{Sushkov2014, Zhang2023}, while the host nuclear spin provides a long-lived memory for enhanced readout~\cite{Bradley2019, Arunkumar2023} and quantum error correction~\cite{Taminiau2014, Abobeih2022}. Together, these complementary roles make electron-nuclear defects versatile building blocks for hybrid NV-based quantum registers.

In addition to the P1, a wide range of electron-nuclear defects can exist in diamond, either formed from nitrogen and hydrogen impurities during material synthesis~\cite{Iakoubovskii2002,Glover2004,Felton2009,Barry2020}, or through ion implantation of a variety of atomic species~\cite{Meijer2005, Tamura2014, Iwasaki2017, Thiering2021}. This diversity offers a rich landscape of spectrally distinct spins through their unique hyperfine couplings, which can be used to overcome the spectral crowding problems associated with controlling the more abundant P1 spin bath~\cite{Degen2021}. Employing these defects as a useful quantum resource requires accurate identification of their hyperfine components and nuclear spin composition, allowing for deterministic control. However, many defects with low natural abundance remain unidentified or poorly characterized due to the limitations of existing electron paramagnetic resonance (EPR) techniques on bulk samples. \red{Defect identification with conventional EPR techniques requires large spin ensembles under strong magnetic fields, which in turn make multiple scans at several crystal orientations necessary to resolve the hyperfine tensor~\cite{Blank2009, Felton2009}}. Recent progress includes NV-based detection of unidentified electron-nuclear defects at the single-spin level~\cite{Cooper2020}, and zero-field electron spin resonance (ESR) of individual P1 centers with kilohertz spectral resolution~\cite{Kong2020}. Still, extending such zero-field techniques to these unidentified defects has yet to be demonstrated. \blue{In addition, these techniques are only sensitive to the magnitudes of the principal hyperfine components. The hyperfine splittings measured at any field orientation, as well as the zero-field transition frequencies, do not reveal the relative sign between components, which is necessary for identification of the defect structure. The relative sign determines how the hyperfine interaction divides into its isotropic (Fermi contact) and anisotropic (dipolar) contributions~\cite{Schweiger2001}. In bulk EPR, this ambiguity is typically resolved with electron-nuclear double resonance (ENDOR), where the nuclear spin transition frequencies depend on the relative sign of the hyperfine and nuclear Zeeman interactions~\cite{Schweiger2001}.} Furthermore, an approach to detect and achieve deterministic control of the defect's host nuclear spin remains an open challenge, given its weak direct coupling to the NV electronic spin. Implementing nuclear spin control through the stronger electron-electron coupling would increase the size of accessible nuclear spin memories~\cite{Stolpe2024} and enable nanoscale magnetic resonance imaging of individual molecules on the surface~\cite{Du2024}.

Here, we develop a self-consistent experimental approach to \blue{characterize} defects at the single-spin level by applying a quantum control protocol called nuclear-electron-electron triple resonance (NEETR). Our approach combines zero-field double electron-electron resonance (ZF-DEER)~\cite{Kong2020} with NEETR to \blue{determine the hyperfine components, including their relative sign, and identify the} host nuclear spin species. We apply this protocol to two unknown electron-nuclear defects \blue{and determine their hyperfine tensors, finding that one is nitrogen-related and the other hydrogen-related. Comparing to density functional theory (DFT) calculations, we assign one defect to a known nitrogen interstitial structure, while the hydrogen-related defect does not match any calculated or previously reported hydrogen center in diamond, suggesting a unique structure.} We further extend the NEETR sequence to demonstrate initialization, coherent control, and millisecond-scale coherence of the hydrogen nuclear spin, highlighting the potential for NV-based hybrid electron-nuclear spin registers.

\begin{figure*}[ht]
	\centering
	\includegraphics[width=6.8in]{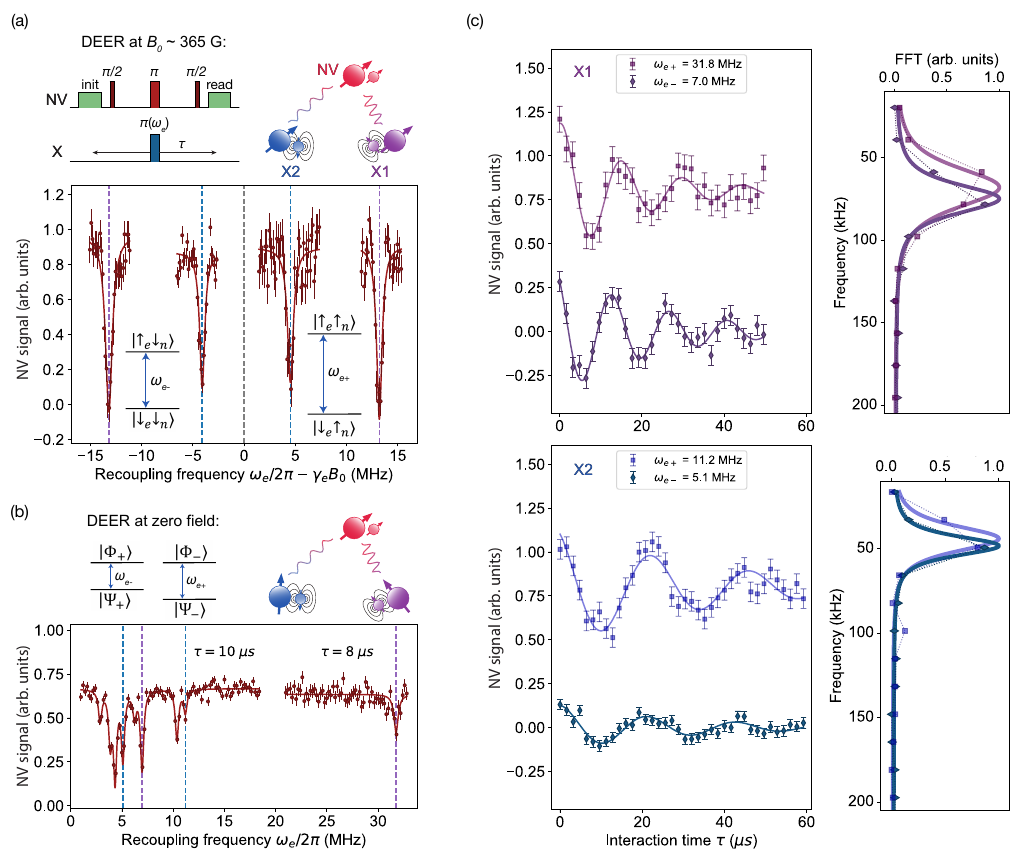}
	\caption{Hyperfine component identification for X1 and X2 defects. (a) DEER performed under an external magnetic field reveals two electron-nuclear spin defects X1 and X2 coupled to the NV electronic spin. \red{Pulse sequence: a spin-echo is applied to the NV probe spin while sweeping the frequency of the recoupling pulse applied to the target spin. The NV spin is initialized and read out with a green laser pulse (see Methods for details on NV signal extraction and data analysis). The (larger) electronic spins of X1 and X2 are quantized along the external field direction, parallel to the NV molecular axis, while the (smaller) nuclear spins are quantized along each defect’s principal hyperfine axis, which is in general misaligned with the field.} For each defect, the two electronic transitions ($\omega_{\mathrm{e}\pm}/2\pi$) are centered around $\gamma_{\mathrm{e}}B_0 \approx  1020$ MHz, with hyperfine splittings $A_1 = 26.4(5)$ and $A_2 =8.6(4)\, \text{MHz}$ for the given magnetic field alignment, confirming $S = 1/2$ electronic and $I = 1/2$ nuclear spins. \red{Each scan for X1 was acquired at $\tau_1\approx 8$ $\mu$s, and $\tau_2\approx 14$ $\mu$s for X2.} (b) The principal hyperfine components are determined by repeating DEER at zero field (ZF-DEER) while sweeping the microwave frequency \red{\recent{over the range of the observable transitions}}. Top schematic: at zero field the electronic spin aligns with the principal hyperfine axis, with transition frequencies $\omega_{\mathrm{e},\pm}/2\pi = \blue{|A_{\parallel}\pm A_{\perp}|/2}$ between Bell states. Two scans were taken with $\tau = 8, 10 \, \mu\text{s}$ based on the different NV-X electronic spin coupling strengths for X1 and X2 \red{and NV decoherence at zero field}. (c) In the presence of additional spins coupled to the NV (see \cref*{sec:smMoredefects}), X1 and X2 transitions are assigned by measuring the coupling strength for each resonance dip. ZF-DEER with swept $\tau$ is performed with $\omega_{\mathrm{e}}$ set to the frequencies at the dashed lines in (b), revealing two pairs of signals with matching coupling strengths ($d_1,d_2 = 70, 47 \, \text{kHz}$), as confirmed by the overlapping FFT spectra on the right. \blue{From the Lorentzian fits in (b), we extract $|A_{\parallel}\pm A_{\perp}|$, which fixes the magnitudes of the principal hyperfine components to 39(1) and 25(1)~MHz for X1, and 16(1) and 6(1)~MHz for X2.}}
	\label{fig:DEER}
\end{figure*}

\section*{Results}

\subsection*{Hyperfine component identification at zero field}

We seek to estimate the components of the hyperfine tensor of two unknown defects through measurements performed on the NV center using pulsed microwave sequences. These hyperfine components describe the interaction between the electron and nuclear spins. They provide a spectral signature that uniquely characterizes the defect's atomic structure, enabling both identification and control of their spin states. We use an isotopically purified diamond sample and create near-surface defects by implanting $^{15}\mathrm{N}$ ions through nanoapertures~\cite{Cooper2020}.
Our experiments focus on one specific NV center that features several nearby $\text{S}=1/2$, $\text{I}=1/2$ electron-nuclear defects. These defects were observed in our previous work~\cite{Cooper2020}, but their structures could not be assigned to any known defects previously detected with EPR. The secular spin Hamiltonian for an electron-nuclear defect in an external magnetic field $B_0$ is
\begin{equation}\label{eq:H0_B}
\begin{split}
    H_0/(2\pi) = \gamma_{\mathrm{e}}B_0S_z + &A_{zx}S_zI_x + A_{zy}S_zI_y \\
    + &A_{zz}S_zI_z \blue{-} \gamma_{\mathrm{n}}B_0I_z,
\end{split}
\end{equation}
where we assume an isotropic  $g$-tensor~\cite{Iakoubovskii2002}, with \blue{$\gamma_{\mathrm{e}}$ ($>0$) and $\gamma_{\mathrm{n}}$ as the electron and nuclear} gyromagnetic ratios, and $A_{zi}$ as the secular components of the defect's hyperfine tensor in the frame with $\mathbf{B} = B_0\hat{\mathbf{z}}$. The eigenstates are labeled in \cref*{fig:DEER}(a), with the electron spin quantized along the external magnetic field direction and the nuclear spin quantized along the sum of the external and hyperfine fields (labeled as $\ket{\uparrow_{\mathrm{n}}}$ and $\ket{\downarrow_{\mathrm{n}}}$ for convenience).

\red{The two defects of interest, which we label X1 and X2, are initially detected through the NV by performing double electron-electron resonance (DEER) spectroscopy~\cite{Degen2021}. The DEER pulse sequence applies a spin-echo to the NV along with a simultaneous $\pi$-pulse to selectively recouple the dipolar interaction between the NV and the nearby defect when driven on resonance. The defect's electronic spin transition frequencies are identified by sweeping the recoupling pulse frequency~(see DEER spectrum in \cref{fig:DEER}a). The interaction time ($\tau$) for the pulse sequence is set according to the dipolar coupling strengths for X1 and X2 as identified in our previous work~\cite{Cooper2020}.} The measurement is performed with the magnetic field aligned along the NV [111] molecular axis, which in general is along a different direction than the principal axis for each defect. While the hyperfine tensor is diagonal in each defect's principal frame, the components in the frame quantized by the electronic spin have a complex dependence on the angular misalignment between the frames (\cref*{sec:smEigen}). As a result, the principal hyperfine components cannot be directly extracted from a single measurement of the hyperfine splitting, that is, the frequency difference between the two electronic transitions. In Ref.~\cite{Cooper2020}, the hyperfine components for these defects were estimated by measuring the hyperfine splittings at various external field orientations. This approach is highly sensitive to miscalibration of the external field as estimated by the NV, leading to a large uncertainty in the predicted hyperfine components, and thus precludes any clear assignment of the defect structure or composition. 

\red{An alternative approach to resolving the hyperfine components for S = 1/2 defects which does not require  careful calibration of the external field or lengthy field scans is to measure the DEER spectrum at zero field (ZF-DEER~\cite{Kong2020})}. \red{This approach is restricted to S = 1/2 defects: for $S > 1/2$, the zero-field splitting typically dominates over the hyperfine interaction and sets the quantization axis instead, so extracting the hyperfine components in this regime would require DEER measurements at several external field orientations, as in Ref.~\cite{Cooper2020}.} \red{For $S = 1/2$ defects at zero field, the spin Hamiltonian is}
\begin{equation}\label{eq:H0_ZF}
    H_0^{\mathrm{ZF}}/(2\pi) = A_{\perp}\left(S_xI_x + S_yI_y\right) + A_{\parallel}S_{z}I_{z},
\end{equation}
where we assume uniaxial symmetry for the defect. The eigenstates of $H_0^{\mathrm{ZF}}$ are the Bell states with electron-nuclear transition frequencies composed of linear combinations of the principal hyperfine components~\cite{Kong2018}. \red{At zero field,    the two transitions which contribute  a measurable phase difference in the NV DEER signal are $\{\ket{\Psi_-} \rightarrow \ket{\Phi_\pm},\ket{\Psi_+} \rightarrow \ket{\Phi_\pm} \}$, with frequencies $\omega_{\mathrm{e},\pm}/(2\pi) = \blue{\left|A_{\parallel}\pm A_{\perp}\right|/2}$ (\cref*{sec:smZFDEER}). For a non-uniaxial defect, the number of transitions increases from two to four. Therefore, we can determine the hyperfine components for a given electron-nuclear defect by identifying these transitions in the ZF-DEER spectrum, as the recoupling frequency is swept over the range covering the maximum number of observable transitions.} \blue{Since these frequencies depend only on $|A_{\parallel}\pm A_{\perp}|$, ZF-DEER determines only the magnitudes of the principal components, and cannot resolve their relative sign nor which sign corresponds to $A_{\parallel}$ or $A_{\perp}$. The same holds for the hyperfine splitting measured with DEER under an external field, which depends only on $A_{\parallel}^2$ and $A_{\perp}^2$ (\cref*{sec:smEigen}).} With more than one defect coupled to the NV, the peaks (dips) in the ZF-DEER spectrum belonging to different defects can be distinguished through their unique dipolar coupling strength to the NV. This requires an additional measurement for each transition by sweeping the interaction time with the recoupling pulse tuned to each transition frequency.

Using the hyperfine splittings from the DEER spectrum as a reference, we apply ZF-DEER and observe several resonances corresponding to the electron-nuclear transitions of multiple defects surrounding the NV~(\cref{fig:DEER}b). To conclusively assign the X1 and X2 transitions in the spectrum, we identify the pairs of resonances corresponding to each defect by matching their coupling strengths to the NV. Varying the interaction time of the ZF-DEER sequence for each resonance, we observe two pairs of signals with matching oscillation frequencies, indicating each pair belongs to a distinct defect with a unique coupling strength to the NV~(\cref{fig:DEER}c). The two different coupling strengths for the X1 and X2 defects are 70 kHz and 47 kHz respectively, \red{consistent with the previously measured values of 66 kHz and 36 kHz~\cite{Cooper2020}, given the different quantization axes at zero field. This corresponds to NV--X distances of $\approx$9 nm (X1) and $\approx$7 nm (X2) as estimated in Ref.~\cite{Cooper2020}}. \blue{From the experimental results, we extract the magnitudes of the principal hyperfine components for both defects: 39(1) and 25(1)~MHz for X1, and 16(1) and 6(1)~MHz for X2.}

\blue{The total number of resonances observed in the ZF-DEER spectrum exceeds the number of transitions expected from two uniaxial defects.  We attribute this to the presence of additional defects rather than additional transitions of X1 and X2 (resulting from $A_{xx} \ne A_{yy}$). This is confirmed by measuring the coupling strengths at these additional resonances with ZF-DEER, which differ from the X1 and X2 couplings or, in one case, from their coherence times (see~\cref*{sec:smMoredefects}). The hyperfine interaction for the X1 and X2 defects is therefore approximately uniaxial, with a maximum deviation of $|A_{xx} - A_{yy}|\blue{\lesssim} 1\,\text{MHz}$ from experimental uncertainty (\blue{\cref*{sec:smResid}}). We also find that the additional resonances appear to fluctuate over the timescale of several days, possibly due to charge-state instability~\cite{Marcks2024}. As a result, we choose to focus our characterization in this work on only the X1 and X2 defects, which are more stable and hence can be harnessed as a reliable quantum resource.} 

We note a discrepancy beyond our experimental uncertainty between these hyperfine components and previous measurements using the swept-external field approach~\cite{Cooper2020}.
We attribute this difference to large uncertainties in the magnetic field calibration required for the previous approach. Comparing \blue{the magnitudes of} these new hyperfine components to other defects in diamond, we do not find any agreement within experimental error with any known S = $1/2$ nitrogen-, hydrogen-, or silicon-related defects measured from EPR experiments~\cite{Zhou1996,Iakoubovskii2002,Atumi2020,Felton2009}. Additionally, we do not find any consistent match with existing DFT results for other defects in diamond with natural concentrations below the EPR detection limit~\cite{Peaker2018}. \red{Thus, in order to search for defect structures with matching hyperfine components through new DFT calculations we first seek to identify the nuclear spin species native to each defect}\blue{, together with the relative sign of the hyperfine components}.

\begin{figure*}[ht]
	\centering
	\includegraphics[width=6.8in]{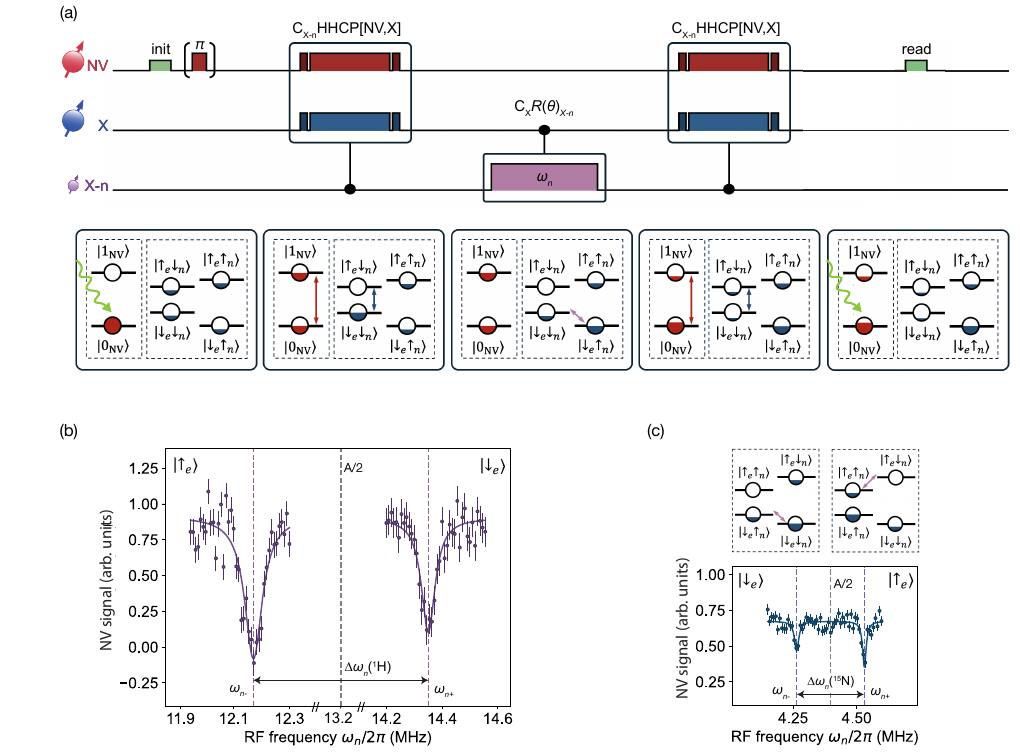}
	\caption{Nuclear spin identification for X1 and X2 defects. (a) Nuclear-electron-electron triple resonance (NEETR) pulse sequence and accompanying spin polarization diagram for \red{a} NV-X system (X, X-n represent the electronic and nuclear spins of the defect, respectively). Polarization is quantified by the difference in filled color between spin states (initially the NV is polarized in $\ket{0}$ while X and X-n are mixed with each state 1/4 occupied)~\red{\cite{Tyryshkin2006}}. The NEETR sequence detects the resonance frequency of the X nuclear spin through changes in \red{the} electronic spin polarization. Hartmann-Hahn Cross Polarization (HHCP) is used to transfer polarization from the NV to X, conditioned on the nuclear spin state by driving a single hyperfine transition. By sweeping the frequency of the RF pulse applied to X-n, we can identify the two nuclear spin transitions when on resonance as the nuclear spin becomes partially polarized. \red{The corresponding change in the electronic spin polarization is mapped to the NV by repeating the conditional HHCP step for readout}. An additional $\pi$-pulse on the NV is applied every other sequence after initialization to reset the X nuclear spin polarization. (b,c) We perform the NEETR experiment on both X1 and X2 defects using the initial magnetic field conditions from DEER~(\cref{fig:DEER}a) by sweeping the \red{frequency of the RF pulse} around $A/2$. (b) The NEETR signal for X1 reveals the defect's nuclear spin resonance frequencies $\omega_{\mathrm{n}\pm}$, with splitting $\Delta \omega_{\mathrm{n}}$ consistent with \blue{a $^{1}$H nuclear spin coupled through opposite-sign hyperfine components (see text)}, supporting the assignment of X1 as hydrogen related. (c) The NEETR signal for X2 features nuclear spin resonances with splitting consistent with the $^{15}\mathrm{N}$ gyromagnetic ratio \blue{for same-sign hyperfine components}, supporting the assignment of X2 as nitrogen related.}
	\label{fig:NVXENDOR}
\end{figure*}

\subsection*{Nuclear spin identification with NEETR}
\red{The hyperfine components identified through ZF-DEER provide a unique spectral signature which enables selective control of each defect's electronic spin. We now introduce the nuclear-electron-electron triple resonance (NEETR) sequence to detect and control each defect's nuclear spin. This sequence combines the conventional pulsed electron-nuclear double resonance (ENDOR) protocol~\cite{Feher1956, Davies1974, Schweiger2001, Tyryshkin2006} with the control tools previously developed for the X1 and X2 electronic spins~\cite{Cooper2020, Ungar2024}. In a typical ENDOR measurement performed on a thermally polarized ensemble of paramagnetic spins, a selective microwave (mw) $\pi$-pulse is applied to the electronic spin for a single hyperfine transition, followed by a selective RF pulse at a swept frequency to transfer polarization to the nuclear spin. The measured signal then corresponds to the loss of the electronic spin polarization when the RF pulse is on resonance with the nuclear spin transition. The NEETR protocol~(\cref{fig:NVXENDOR}a) implements this scheme for a defect \red{whose nuclear spin has} negligible direct coupling to the NV by instead using the stronger electronic spin coupling between the defect and the NV to probe the nuclear spin transition.}

\red{The sequence consists of three steps: (1) polarization of the X electronic spin, (2) ENDOR on the X electron-nuclear spin system, and (3) readout of the X electronic spin polarization. For steps (1) and (3), the X electronic spin is initialized and read out by applying the Hartmann-Hahn Cross Polarization (HHCP) sequence~\cite{Hartmann1962} through simultaneous resonant driving of the NV and X spins with matched Rabi frequencies ($\Omega_{\text{NV}} = \Omega_{\text{X}}$). Both the electronic spin polarization and selective mw pulse of the conventional ENDOR protocol are implemented \red{within} a single step by applying HHCP on a single hyperfine transition (C$_{\text{X-n}}$HHCP[NV,X]). By addressing a single hyperfine transition, the maximum electronic spin polarization is 50\% of the initial NV polarization. This partial polarization is transferred to the nuclear spin with the conditional RF $\pi$-pulse, completing the ENDOR sequence. The corresponding loss of electronic spin polarization from the addressed hyperfine manifold is measured by repeating the conditional HHCP step followed by fluorescence readout on the NV. We apply an alternating $\pi$-pulse on the NV which cycles the X nuclear spin polarization between $\ket{\uparrow_{\mathrm{n}}}$ and $\ket{\downarrow_{\mathrm{n}}}$ to avoid long signal recovery times resulting from the nuclear spin-lattice relaxation time $T_{1,\mathrm{n}}$~\cite{Tyryshkin2006}. This sequence provides a solution for detecting and controlling distant nuclear spins with negligible coupling to the NV by leveraging the stronger electronic spin coupling between the NV and the defect. In addition to defect identification, this approach can be extended to addressing reporter electronic spins for performing nuclear magnetic resonance of single molecules on the surface~\cite{Sushkov2014, Zhang2023}.}

\red{We now} identify the native nuclear spin species for the X1 and X2 defects by applying the NEETR protocol at the same external field condition as in the initial DEER measurement~(\cref{fig:DEER}a). \blue{For the $\text{I} = 1/2$ nuclear spin isotopes of X1 and X2, the two nuclear spin transition frequencies are centered about half the hyperfine splitting, $A/2$, and separated by (\cref*{sec:smEigen})
\begin{equation}\label{eq:dwn}
\Delta\omega_{\mathrm{n}}/(2\pi) \approx 2\left|\gamma_{\mathrm{n}}B_0\frac{A_{zz}}{A} + \frac{A_{\perp}^2A_{\parallel}}{4\gamma_{\mathrm{e}}B_0A}\right|,
\end{equation}
where $A_{zz} = A_{\parallel}\cos^2\vartheta + A_{\perp}\sin^2\vartheta$ is the secular hyperfine component along the field direction, $A = (A_{\parallel}^2\cos^2\vartheta + A_{\perp}^2\sin^2\vartheta)^{1/2}$ is the hyperfine splitting, $\vartheta$ is the angle between the defect's principal hyperfine axis and $B_0$, and the second term is a small correction from the non-secular hyperfine terms. In \cref*{sec:smEigen} we show that for same-sign components, $A_{zz}\approx A$ and $\Delta\omega_{\mathrm{n}}/(2\pi)\approx 2\gamma_{\mathrm{n}}B_0$ for any orientation, so that the splitting directly identifies the nuclear spin species through its gyromagnetic ratio. For opposite-sign components, $|A_{zz}|/A$ can range from zero to one depending on the angle of the principal hyperfine axis. In this case, the hyperfine splitting measured with DEER fixes $\vartheta$, and the gyromagnetic ratio can be determined or constrained through the hyperfine components in combination with the nuclear splitting. In addition to nuclear spin identification, the NEETR splitting also resolves the relative sign of the hyperfine components, which is inaccessible to DEER and ZF-DEER alone.} 

\red{Applying the NEETR sequence, we measure the nuclear spin resonance spectra for X1 and X2 with the RF pulse frequency scanned around their respective hyperfine splittings ($A/2$), and identify the two resonances for each defect which correspond to the nuclear spin transitions in the different electronic spin manifolds~(\cref{fig:NVXENDOR}b,c).} \blue{For X1, the measured splitting of 2.2~MHz between the nuclear spin resonances excludes all $\text{I}=1/2$ isotopes plausible as impurities in diamond other than $^1$H and $^{19}$F, for either relative sign of the hyperfine components (\cref*{sec:smResid}). For these two isotopes, same-sign hyperfine components would give $\Delta\omega_{\mathrm{n}}/(2\pi)\approx 2\gamma_{\mathrm{n}}B_0\approx 3$~MHz at any defect orientation, so the hyperfine components of X1 must have opposite signs. This indicates that the hyperfine interaction is predominantly dipolar rather than nearly isotropic, as it would be for same-sign components. A global fit of all measured transition frequencies from DEER, ZF-DEER, and NEETR then yields two possible solutions for both $^1$H and $^{19}$F isotopes up to an overall sign (\cref*{sec:smResid}): $(A_{\parallel,1}, A_{\perp,1}) = (+39(1), -26(1))$~MHz or $(+23(1), -38(1))$~MHz.} \red{Given the
large concentration of hydrogen impurities introduced during CVD growth, we believe this defect is most likely hydrogen-related~\cite{Cann2009, Barry2020, Glover2004}. We note that our sample was annealed at 800$^\circ$C (see Methods), which is below the threshold at which hydrogen can be annealed out~\cite{Mizuochi2004}, and may assist in the formation of hydrogen-vacancy complexes~\cite{Iakoubovskii2002}.} \blue{For X2, the measured splitting of 0.27~MHz agrees with a $^{15}\text{N}$ nuclear spin with same-sign hyperfine components, for which $\Delta\omega_{\mathrm{n}}/(2\pi)\approx 2|\gamma_{\mathrm{n}}|B_0\approx 0.3$~MHz, and the global fit yields $(A_{\parallel,2}, A_{\perp,2}) = (16(1), 6(1))$~MHz, confirming X2 as a nitrogen-related defect. An opposite-sign solution is only slightly disfavored and cannot be excluded, in which case the splitting alone would not constrain the nuclear species (\cref*{sec:smResid}).} \blue{\recent{However, the} presence of a nitrogen-related defect} is consistent with our expectation as the sample was implanted with $^{15}\text{N}$ ions and the NV conversion efficiency is on the order of a few percent~\cite{Schwartz2011}.

\blue{The results of DEER, ZF-DEER, and NEETR are self-consistent as the resulting signed hyperfine tensors reproduce all measured transition frequencies of both defects within experimental uncertainty (\cref*{sec:smResid}). Together with the identified nuclear spin species, these measurements provide a reference for the DFT calculations below. We note that from a single NEETR spectrum, the nuclear spin species is in general identified only up to a small discrete set of candidates. Here the assignment relies on only one candidate being chemically plausible out of the possible set for each defect. Additional DEER and NEETR measurements at a second field orientation would provide further constraints to lift this ambiguity for general defects.}

\subsection*{\blue{Comparison with calculated and known defect structures}}

\blue{We perform DFT calculations to search for atomic structures matching the X1 and X2 defects, constrained by the measurements from DEER and NEETR: an $\text{S}=1/2$ electronic spin, a single hydrogen or nitrogen nuclear spin, and the signed hyperfine components found above. 
We restrict the search to defects with a single hydrogen or nitrogen atom, since complexes with several hydrogen atoms are less likely given the fabrication process. A single hydrogen or nitrogen atom gives an odd number of electrons in the neutral charge state, as required for $S = 1/2$, and we therefore consider charge-neutral defects (doubly charged defects being less likely).
Following the \textit{ab initio} methods developed in Ref.~\cite{Tang2023}, we calculate the hyperfine tensors of charge-neutral hydrogen and nitrogen defect structures, considering both interstitial and substitutional defects with zero to two vacancies. \blue{\recent{We did not find a match for X1 after performing calculations for every symmetry-distinct configuration of a single hydrogen atom with up to two nearby vacancies}, as well as hydrogen complexes with a carbon self-interstitial} (see Methods and \cref*{sec:smDFT}). For X2, the closest match is the nitrogen interstitial N$_\text{I}^0$, with calculated principal components of 9, 11, and 13~MHz using the PBE functional, all of the same sign, in agreement with the measured components. This structure corresponds to the WAR9 defect previously characterized with EPR and DFT~\cite{Atumi2013}. We further find that the neutral charge state of WAR9 is stable over a range of Fermi levels that overlaps with the stability range of NV$^-$ (\cref*{sec:smIonization}). Given this agreement and the $^{15}$N implantation of the sample, we did not search the nitrogen-related structures as extensively (\cref*{sec:smDFT}).}

\blue{For \recent{the calculated hydrogen defects, most structures retain several carbon dangling bonds, for which the calculated tensor depends on the spin configuration reached in the calculation. For these structures, we therefore evaluated the hyperfine tensor for the spin-$1/2$ states of the dangling-bond electrons, rather than for a single calculated spin configuration. For 68 out of the 95} calculated structures, \recent{no spin-$1/2$ state reproduces} X1 \recent{within the accuracy of the calculation, although for 16 of them only some of the spin-$1/2$ states could be tested (\cref*{sec:smDFT}). The closest case, a structure 5.6~eV above the lowest-energy structure of its composition, comes within 2~MHz of the principal values of X1, but its principal axis lies $7^\circ$ outside the orientation range fitted from the measured frequencies. For the remaining structures, which are not energetically favorable (at least 2.8~eV above the lowest-energy structure of their composition), the analysis is inconclusive, could not be applied, or was limited to a single spin configuration} (\cref*{sec:smDFT}). \recent{The calculations do not cover complexes with more than one hydrogen atom, larger vacancy clusters, or other impurities, so a} vacancy--hydrogen complex \recent{outside our calculated set remains possible} as the origin of X1.}

\blue{We further compare X1 to the hydrogen-related centers in diamond characterized with EPR and to previously calculated hydrogen-related defect structures~\cite{Zhou1996, Glover2003, Iakoubovskii2002, Cann2009, Hartland2014, Peaker2018, Thiering2016} (\cref*{sec:smDFT}), and find no structure that matches the X1 measurements. Based on NEETR, the X1 hyperfine tensor features a large dipolar contribution, $T = (A_{\parallel}-A_{\perp})/3\approx 21$~MHz. This dipolar contribution exceeds that of all previously reported hydrogen-related centers with signed tensors ($T\lesssim 18$~MHz), suggesting a proton located closer to localized electronic spin density than in previously characterized hydrogen defects. The closest known center, WAR2~\cite{Cann2009}, shares the opposite-sign pattern of X1, but its strongly non-uniaxial tensor predicts four ZF-DEER transitions that are inconsistent with the X1 spectrum, and its predicted NEETR splitting is inconsistent with the measured value (\cref*{sec:smDFT}). We therefore conclude that X1 is a hydrogen-related defect distinct from all previously characterized hydrogen centers in diamond. The orientation of its principal axis fitted from the DEER and NEETR frequencies ($\vartheta\approx 78^\circ$ or $24^\circ$) further excludes trigonal centers with a $\langle 111\rangle$ axis, as well as $\langle 110\rangle$- or $\langle 100\rangle$-oriented axes, indicating a low-symmetry structure. Identifying X1 therefore calls for a dedicated computational study of a wider range of low-symmetry hydrogen-related structures.}

\begin{figure*}[ht]
	\centering
	\includegraphics[width=6.8in]{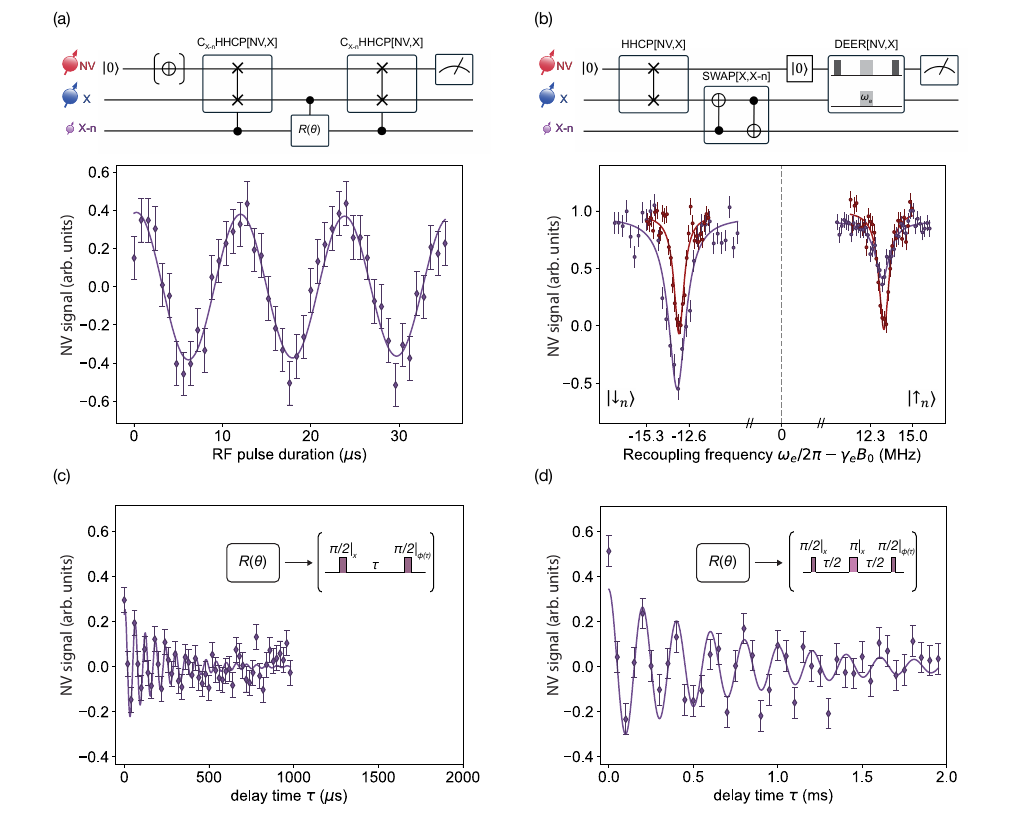}
	\caption{Universal control and coherence of the X1 hydrogen nuclear spin. (a) Applying the NEETR sequence~(see \cref{fig:NVXENDOR}a) with the RF pulse tuned to the X1 nuclear spin transition $\omega_{\mathrm{n}+}$ to achieve coherent control of the nuclear spin qubit. The signal shows several Rabi oscillations of the hydrogen nuclear spin with negligible decay. \red{Top panel: quantum circuit representation of the NEETR sequence.} The conditional HHCP steps implement an $i$SWAP operation between the NV and X electronic spins for a single hyperfine manifold. The selective nuclear spin drive implements a controlled rotation, conditioned on the electronic spin state. The $i$SWAP is repeated for readout of the nuclear spin state \red{through the NV fluorescence}. (b) Demonstrating \red{near-maximal} initialization of the hydrogen nuclear spin by performing sequential polarization transfer over the NV-X-Xn system. Both X hyperfine transitions are driven during HHCP for a full $i$SWAP between NV and X, and two conditional gates are applied to X for the electron-nuclear SWAP. We probe the nuclear spin polarization by measuring the DEER spectrum over the two hyperfine transitions. The amplitudes for each dip after the initialization sequence (purple) grow or shrink compared to the reference signal without initialization (red), confirming nuclear spin polarization with a fidelity of \red{0.57(7)} (\cref*{sec:smFidelity}). (c,d) Coherence measurements of the hydrogen nuclear spin qubit using the NEETR sequence with Ramsey and spin-echo evolution applied during the conditional $R(\theta)$ block (see insets). The last $\pi/2$-pulse has a modulated phase, with $\phi =  2\pi f_{\text{mod}}\tau$, to improve the estimate of the decay constant using a fit to an exponentially decaying cosine function. (c) Measuring the nuclear spin dephasing time with phase-modulated Ramsey at $f_{\text{mod}}= 20 \, \text{kHz}$, we find $T_2^{*} = 250(60) \,\mu\text{s}$. (d) Measuring the nuclear spin coherence time using phase-modulated spin-echo at $f_{\text{mod}} = 5 \,\text{kHz}$, we find $T_2 = 1.0(3)\, \text{ms}$.}
	\label{fig:control}
\end{figure*}

\subsection*{\red{Control and coherence of the hydrogen nuclear spin}}

The identification of previously unknown defects provides a route to individually controlled electron-nuclear spin registers for quantum sensing and information processing. Each \red{unique} defect hosts a distinct electronic spin transition — useful as a spectrally addressable sensing qubit — together with a host nuclear spin that can serve as a long-lived quantum memory. This dual functionality enables selective control with minimal spectral overlap from other environmental impurities such as P1 centers. Hydrogen-related defects are particularly promising because of their lower concentrations relative to nitrogen, which improves distinguishability through NV-electron coupling. Furthermore, access to defects along electronically-coupled spin chains~\cite{Ungar2024} can reduce the number of distinct defect species required to build larger hybrid registers. \red{While the hydrogen-related defect found in this work likely formed from hydrogen incorporated during CVD growth, hydrogen-related defects can be introduced in a more controlled way through the proper growth and annealing conditions~\cite{Glover2004}, or by hydrogen ion implantation, as for the analogous vacancy--hydrogen complex in silicon~\cite{BechNielsen1997}.} The key requirements for harnessing electron-nuclear defects as quantum memories are initialization, coherent control, and readout of the defect’s nuclear spin qubit. Using the calibrated nuclear spin transitions of the X1 defect~(\cref{fig:NVXENDOR}b), we extend the NEETR sequence to achieve universal control \red{of} the nuclear spin \red{qubit and demonstrate its potential} as a robust quantum memory.

For unitary control, we apply the NEETR sequence \red{introduced in~\cref{fig:NVXENDOR} with the RF pulse (C$_\text{X}R(\theta)_{\text{X-n}}$)} tuned to either of the $\omega_{\mathrm{n}\pm}$ frequencies, which implements a selective nuclear spin rotation conditional on the electronic spin state. By varying the RF pulse duration we observe multiple Rabi oscillations of the hydrogen nuclear spin~(\cref{fig:control}a). We calibrate the $\pi$-pulse duration based on the measured Rabi frequency in order to implement the conditional nuclear spin gate \red{used in the} nuclear \red{polarization protocol}, as shown below. The fast nuclear Rabi frequency of hydrogen relative to nitrogen, in combination with the long nuclear coherence times observed below, will help to enable efficient qubit control for quantum memory protocols \red{in an electron-nuclear spin register}.

We achieve \red{deterministic} nuclear spin initialization by performing concatenated polarization transfer from the NV electronic spin to the X1 hydrogen nuclear spin. \red{Building on the NEETR sequence, we introduce a modified protocol designed to achieve near-maximal nuclear spin polarization (see sequence in \cref{fig:control}b). We first apply HHCP to implement a full $i$SWAP between the NV and X1 electronic spins \red{by driving both hyperfine transitions of X1}. This is followed by a selective microwave $\pi$-pulse on the $\omega_{\mathrm{e},+}$ transition, and then a selective RF $\pi$-pulse on the $\omega_{\mathrm{n},+}$ transition. The two selective (conditional) gates applied sequentially constitute a SWAP operation between the X1 electronic and nuclear spins.} \red{We confirm} nuclear spin polarization of the hydrogen defect by comparing the DEER spectra \red{with and without the initialization sequence~(\cref{fig:control}b)}. We observe the two hyperfine dips grow and shrink proportionally after performing the initialization sequence, consistent with the nuclear spin being polarized in the $\ket{\downarrow_{\mathrm{n}}}$ state. From the integrated area of each peak, we estimate the polarization of the hydrogen nuclear spin to be approximately 60\%, which is slightly smaller than the maximum expected polarization of 70\% when accounting for the initial NV polarization \red{fidelity} and experimental errors in HHCP (\cref*{sec:smFidelity}). \red{Repeating the initialization sequence up to three times increases the polarization only slightly, within the measurement uncertainty (\cref*{sec:smFidelity}).} With NEETR, the nuclear spin polarization and readout steps rely on the stronger electronic spin coupling between the NV and X defect to mediate control. \red{This provides a route to} constructing larger electron-nuclear registers with access to long-lived nuclear spin quantum memories, as demonstrated below.

We highlight the potential of the hydrogen defect's nuclear spin qubit as a quantum memory by demonstrating millisecond-scale coherence at room temperature. We apply both nuclear Ramsey and spin-echo protocols through the selective nuclear spin control of the NEETR sequence~(\cref{fig:NVXENDOR}a). In order to extract a more precise estimate of the decay constant, we modulate the phase of the second $\pi/2$-pulse and fit the data to an exponentially decaying cosine function (see Methods). Since we measure the nuclear spin state through the defect's electronic spin population, the total signal decay includes decoherence of the nuclear spin as well as relaxation of the electronic spin. We account for this electronic spin induced decay by including an additional time constant of $T_1^{\mathrm{e}}$ in our fits (\cref*{sec:smT1e}). Using the Ramsey sequence, we measure the dephasing time for the hydrogen nuclear spin, finding $T_2^{*}(^1\text{H}) = 250(60)\,\mu\text{s}$~(\cref{fig:control}c). The spin-echo sequence  measures a coherence time of $T_2(^1\text{H}) = 1.0(3) \,\text{ms}$~(\cref{fig:control}d). The coherence measurements for the X2 nitrogen nuclear spin are included in \cref*{sec:smT2n}, where we find $T_2^{*}(^{15}\text{N}) = 150(30)\,\mu\text{s}$ and $T_2(^{15}\text{N}) = 1.0(3) \, \text{ms}$. The significant improvement from $T_2^{*}$ to $T_2$ by refocusing with the echo suggests the dephasing is dominated by interactions from slowly varying magnetic fields, and coherence times can be further improved with dynamical decoupling~\cite{Chen2018}. With \red{a} nuclear spin coherence \red{time} approximately twenty times longer than the electronic spin coherence ($T_2^{\mathrm{e}}\approx50 \, \mu\text{s}$~\cite{Ungar2024}), the long-lived nuclear spin qubit can be used as an effective ancilla for enhanced sensing schemes such as quantum non-demolition (QND) repetitive readout~\cite{Neumann2010} and error correction~\cite{Taminiau2014}. Additionally, the \red{nuclear spins of both X defects} exhibit coherence times similar in magnitude to the native nitrogen nuclear spin of the NV center~\cite{Chen2018}, making both defects promising candidates for electron-nuclear registers with selective control.

\section*{Discussion}
In this work, we demonstrate how the NEETR protocol provides a combined characterization and control solution for harnessing previously unknown electron-nuclear spin defects at room temperature. The NEETR sequence leverages the stronger electron-electron spin coupling between the NV and a nearby defect, allowing the NV to probe and read out a distant nuclear spin with negligible direct coupling. \blue{Combining NEETR with zero-field double electron-electron resonance (ZF-DEER) measurements, we determine the signed hyperfine tensors and nuclear spin species of two unknown defects. Using DFT calculations, we assign X2 to the known nitrogen interstitial WAR9, while X1 is a hydrogen-related defect that does not match any calculated or previously reported structure, motivating further \textit{ab initio} searches. The relative sign determined with NEETR sets the isotropic and dipolar contributions of the hyperfine interaction that are compared with calculated and known structures. For X1, the DEER and ZF-DEER spectra alone would be consistent with a nearly isotropic interaction for same-sign components, whereas the NEETR splitting establishes opposite signs and identifies the interaction as predominantly dipolar in nature.} In combination, the experimental sequences and \textit{ab initio} calculations provide a self-consistent protocol to \blue{constrain} the atomic structure of previously unknown \red{$S = 1/2$} defects in diamond. \red{Our approach can be extended to other ion-implanted diamond samples as well as to color centers in other material hosts such as Si and SiC~\cite{Widmann2015, Bergeron2020}, and in the future to 2D materials such as hBN with further improvements in coherence~\cite{Guo2023, Gong2023, Gong2024}. Extending our identification protocol to a given material platform requires a decoherence rate comparable to or slower than the electron-electron coupling strength between the sensing and target spins ($d\approx 50$--$70$~kHz in this work), i.e. $T_2 \gtrsim 10$--$20\,\mu\text{s}$. While single defects in hBN currently reach coherence times of only $\approx 2.5\, \mu\text{s}$~\cite{Guo2023}, isotope engineering and dynamical decoupling can together be applied to improve coherence by several-fold~\cite{Gong2024, Gong2023}.}

Using NEETR, we achieve initialization, coherent control, and millisecond-scale coherence of the hydrogen defect’s nuclear spin, highlighting its potential as a robust quantum memory.  The \blue{characterization of this hydrogen-related defect}, together with the control of both its electron and nuclear spins, sets the stage for integrating other hydrogen-related defects into hybrid quantum registers.  Applying this \red{defect identification approach} in parallel with new techniques for probing and controlling larger networks of nuclear spins around individual electronic spin nodes~\cite{Stolpe2024, Goldblatt2024} will help to enable future testbeds for error-corrected sensing~\cite{Goldstein2011} and solid-state quantum repeaters~\cite{Pompili2021, Bradley2022}.

\section*{Methods}\label{sec:methods}
\subsection*{Sample fabrication}
The diamond sample used was a single crystal chemical vapor deposition (CVD) diamond from Element Six with a 100 $\mu$m-thick layer of isotopically enriched 99.999\% $^{12}$C grown on top of a 300 $\mu$m-thick electronic grade single crystal diamond substrate. The sample was implanted with $^{15}\mathrm{N}$ ions through 30 nm diameter circular apertures at an energy of 14 keV and dose of $10^{13}$ $\mathrm{cm^{-2}}$, resulting in NV centers approximately 20 nm from the surface. \red{Following implantation, the sample was annealed at 800$^\circ$C for 4 hours in vacuum.} For further details on diamond sample preparation see Supplementary Material of Ref.~\cite{Cooper2020}. 

\subsection*{Hardware setup \red{and microwave control}}
NV center measurements were performed using a home-built confocal microscope setup. A 520 nm pulsed diode laser (LABS electronics DLnsec) was used for green excitation to perform NV polarization and readout. The red fluorescence is passed through a dichroic mirror and detected using a single photon counting module (Perkin Elmer SPCM-AQRH-15-FC) collected through a 100x, 1.3 NA objective (Nikon CFI Plan Fluor). The bias magnetic field is produced from an N52 neodymium 1 inch cube magnet, aligned to the [111] NV crystallographic axis. 

We use a Quantum Machines OPX+ to generate microwave and RF pulses. For NV and X electronic spin control the frequency of the microwave pulses is up-converted with an IQ mixer using an SRS signal generator (SG384). The NV and X microwave signals are amplified using RF power amplifiers Mini-Circuits ZHL-16W-43-S+ and ZHL-20W-13+, respectively. The RF pulses for nuclear spin control are amplified using a Mini-Circuits LZY-22+ RF power amplifier. We use a custom-designed two-channel Cu coplanar waveguide (CPW) for microwave and RF delivery, with electronic and nuclear spin tones sent through the different channels. The NV microwave control is applied using a calibrated 1 MHz Rabi frequency on resonance with a single hyperfine transition. \red{For the DEER and ZF-DEER measurements in~\cref{fig:DEER}a,b the recoupling pulse duration is fixed to 1 $\mu$s. For each transition in DEER, we independently calibrate the recoupling pulse amplitude corresponding to a $\pi$-rotation on the X spin in order to maximize the signal contrast for each DEER peak (dip). For the two scans in ZF-DEER, the pulse amplitude is fixed and calibrated only for a single transition. Due to the frequency dependent transmission of the coplanar waveguide, this leads to different signal contrasts for each transition in the spectrum.}

\subsection*{Measurement acquisition and uncertainty}\label{subsec:error}
We process the NV fluorescence data into the ``NV signal’’ traces displayed in all main text figures by performing a differential measurement for a given microwave control sequence. The differential sequence consists of four measurement traces of the NV photon counts: (1) microwave sequence followed by optical readout with a laser pulse for trace $R_i^{+}$ (2) microwave sequence followed by a $\pi$-pulse on the NV, then readout for trace $R_i^{-}$ (3) repolarization into $m_{\mathrm{s}} = 0$ followed by readout for trace $R_i^{0}$ (4) repolarization followed by a $\pi$-pulse to $m_{\mathrm{s}} = -1$, then readout for trace $R_i^{1}$. Here the subscript ‘‘$i$’’ denotes the index of the independent variable in the control sequence (e.g. time step or frequency value).  To achieve sufficient signal to noise ratio (SNR) we average over approximately $10^5$ to $10^6$ measurements for each trace, to obtain $\{\bar{R}_i^{+}, \bar{R}_i^{-}, \bar{R}_i^{0}, \bar{R}_i^{1}\}$. The averaged differential and normalized ‘‘NV signal’’ trace is then $y_i = (\bar{R}_i^{+} - \bar{R}_{i}^{-})/(\langle \bar{R}_{i}^0\rangle - \langle \bar{R}_{i}^{1}\rangle)$, where $\langle \rangle$ is the average over the independent variable. The associated error bars for the ‘‘NV signal’’ traces are reported as plus or minus one standard deviation, calculated from the combined error of the reference measurements $\sigma_y = \sqrt{(\sigma^0)^2 + (\sigma^1)^2}/(\langle R_{i}^0\rangle - \langle R_{i}^{1}\rangle)$.  

\subsection*{Data analysis and fitting}\label{subsec:fitting}
Each resonance in the DEER spectrum~(\cref{fig:DEER}a, \cref{fig:control}b) and the NEETR spectrum~(\cref{fig:NVXENDOR}b) is fit to a single-Lorentzian function with four free parameters $\{a,b,\gamma,f\}$: $y(x) =\frac{a\gamma^2}{\gamma^2 + \left(x - f\right)^2} + b$. The frequency uncertainty for each resonance is reported as the half width at half maximum (HWHM), equal to $\gamma$. The data in the ZF-DEER spectrum~(\cref{fig:DEER}b) and NEETR spectrum~(\cref{fig:NVXENDOR}c) are fit to a multi-Lorentzian function defined by $y(x) =\sum_{i}\frac{a_i\gamma^2}{\gamma^2 + \left(x - f_i\right)^2} + b$. The ZF-DEER data for $\tau = 10 \,\mu\text{s}$ includes eight peaks and the data for $\tau = 8 \, \mu\text{s}$ includes one peak. The NEETR data is fit to a double-Lorentzian. The ZF-DEER time traces to measure the coupling strengths~(\cref{fig:DEER}c) are fit to an exponentially decaying cosine function with an additional linear baseline offset to account for background decay: $ y(t) = a\cos(\omega t +\phi)e^{ -t/T } + bt + c$. The additional background decay is likely due to electronic noise from the low-frequency recoupling pulse. The linear offsets are subtracted off from both the fit and data for clear comparison of the coupling frequency between measurements. We compute the Fast Fourier Transform (FFT) for the background corrected data and display the power spectral density normalized to the peak power along with their single-Lorentzian fits using interpolated data. The nuclear-Rabi measurement for the X1 hydrogen spin~(\cref{fig:control}a) is fit to $y(t) = a\cos(\omega t +\phi)e^{ -t/T } + bt + c$, and the displayed data and fits have been background subtracted. The coherence measurements~(\cref{fig:control}c,d) are first background subtracted and then fit to exponentially decaying cosine functions with a fixed additional decay term to account for electronic spin relaxation: $y(t) = a\cos(\omega t +\phi)e^{[ -t/T_2^{(*)} - 3t/(2T_1^{\mathrm{e}})]}$, where $T_2^*$ is for Ramsey and $T_2$ is for spin-echo (see \cref*{sec:smT1e}). \red{The phase modulation of the final $\pi$/2-pulse adds a frequency shift to the coherence signal in order to remove any low frequency noise components, similar to time-proportional phase incrementation in NMR~\cite{Marion1983} and to the artificial-detuning approach used in pulsed EPR and quantum sensing~\cite{Degen2017, Schweiger2001}. The frequency of the fit is then matched to the modulation frequency to estimate the signal amplitude, akin to lock-in detection where the filtering step is applied during data post-processing.}

\red{\subsection*{DFT calculations}
The DFT calculations of hydrogen and nitrogen defects are implemented using the Vienna Ab initio Simulation Package (VASP) with the projector-augmented wave (PAW) basis set. Different defect structures are first screened using the generalized gradient approximation with the Perdew-Burke-Ernzerhof (PBE) exchange-correlation functional in $3\times3\times3$ supercells of the diamond cubic lattice (216 lattice sites, $10.72$ \AA), with a 400 eV cutoff energy and $\Gamma$-only k-point sampling.
\blue{Each structure is relaxed} with atomic forces converged to less than 0.01 eV/\AA\ and electronic energies converged to $10^{-5}$ eV. \blue{The hydrogen-related structures are calculated with VASP 6.6.1 and relaxed without symmetry constraints, starting from atomic positions randomly displaced by 0.02~\AA, with the total spin fixed to $S = 1/2$. \recent{Their} hyperfine matrices are evaluated with single-point calculations using the HSE06 hybrid functional on the PBE geometries\recent{, for 93 of the 95 structures, and with PBE for the remaining two higher-energy structures. The lowest-energy structure of each composition with several dangling bonds and the structure closest to X1 are also relaxed with HSE06}. HSE06 reproduces the measured $^1$H hyperfine tensor\recent{s} of the N$_2$VH$^0$ center\recent{, with a proton next to a single carbon dangling bond, to} within 1~MHz\recent{, and of the NVH$^-$ center, with one unpaired electron shared by two dangling bonds, to within 2.1~MHz} (\cref*{sec:smDFT}). For \recent{structures with several carbon dangling bonds, whose calculated hyperfine tensor depends on the spin configuration, we additionally evaluate the $^1$H tensor for all $S = 1/2$ states of the dangling-bond electrons. For} nitrogen, PBE and HSE06 reproduce the measured $^{14}$N hyperfine components of NV$^-$ to within $\approx 20\%$. The nitrogen-related structures are relaxed within the symmetry of their initial configuration with the total spin unconstrained. The closest-matching nitrogen structure, WAR9, is recalculated with the settings used for the hydrogen-related structures above, and its hyperfine matrix is evaluated with both PBE and HSE06.} The hyperfine matrix is evaluated from the converged electronic structure as a sum of the isotropic Fermi contact interaction and the anisotropic magnetic dipolar interaction. Defect candidates are matched to the experimental hyperfine components using the distance metric $d_A = \sqrt{(\langle A_{\perp}^{\mathrm{c}}\rangle - A_{\perp})^2 + (A_\parallel^{\mathrm{c}}-A_{\parallel})^2}$ in the three-dimensional principal-value space of the calculated hyperfine matrix, where $\langle A_{\perp}^{\mathrm{c}}\rangle = (A_1^{\mathrm{c}} + A_2^{\mathrm{c}})/2$ and $A_\parallel^{\mathrm{c}} = A_3^{\mathrm{c}}$ \blue{are signed principal values. Since only the relative sign of the experimental components is determined, the metric is evaluated for both overall signs of the experimental tensor} (see \cref*{sec:smDFT} for further details).}

\subsection*{\blue{Use of AI tools}}
\blue{Claude (Anthropic), a large language model, was used to assist with parts of this work: deriving the dependence of $A_{zz}/A$ on the relative sign of the hyperfine components and the correction to the nuclear spin frequency splitting from the non-secular hyperfine terms (\cref*{sec:smEigen}), writing the code for the exact diagonalization and global fits of the measured transition frequencies (\cref*{sec:smResid}), surveying the literature on hydrogen-related defects in diamond (\cref*{sec:smDFT}), running and analyzing the final round of DFT calculations for the hydrogen-related structures, and drafting and editing parts of the text. The authors reviewed all analyses and take full responsibility for the content of the manuscript.}

\section*{Data availability}
\blue{The data underlying the figures of this research article, together with the DFT input files, relaxed structures, calculation outputs, and calculated hyperfine tensors for the defect structures in \cref*{sec:smDFT} and the charge-state energies in \cref*{sec:smIonization}, are available online through \url{https://doi.org/10.5281/zenodo.23782606}. \recent{The calculated hyperfine tensors of all hydrogen-related structures are also provided as Supplementary Data~1.}}

\section*{\blue{Code availability}}
\blue{The code for the exact diagonalization and global fits of the measured transition frequencies (\cref*{sec:smEigen,sec:smResid}) and the scripts used to set up and analyze the DFT calculations are available in the same record, \url{https://doi.org/10.5281/zenodo.23782606}. The DFT calculations used the commercial VASP code (version 6.6.1), which is not included.}

\nolinenumbers
\putbib[ENDOR_fixed]

\begin{acknowledgments}
We gratefully acknowledge helpful conversations with Garrett Heller. We wish to thank the T.J. Rodgers RLE Laboratory, Massachusetts Institute of Technology (MIT) for providing equipment and facilities support. This material is based upon work in part supported by the National Science Foundation (NSF) under Grant No. PHY1734011 and Graduate Research Fellowship Program under Grant No. 4000181759. The work of A.S. was supported by the NSF Grant No. PHY1915218. The work of A.C. was supported by the Canada First Research Excellence Fund (CFREF). The work of B.X. was supported by the A*STAR International Fellowship. \blue{The authors acknowledge the MIT Office of Research Computing and Data for providing high performance computing resources that have contributed to the research results reported within this paper.}
\end{acknowledgments}

\section*{Author contributions}
A.U. designed and carried out the experiments, led the theoretical modeling for the experimental results, analyzed the data, \blue{performed the final DFT calculations,} and wrote the manuscript. H.T. performed the \blue{initial} DFT calculations \blue{and the charge stability analysis, helped analyze the final round of calculations,} and contributed to the manuscript.
 A.C. designed the experimental setup, performed preliminary measurements characterizing the defects, and helped conceive the idea for the electron-nuclear register. A.C., A.S., and B.L. assisted with the experimental setup. B.X., A.S., and P.C. contributed to the theoretical derivations. A.C., H.T., A.S., B.X., and P.C. helped with manuscript preparation. P.C. supervised the project.

\section*{Competing interests}
The authors declare no competing interests.
\clearpage
\end{bibunit}

\widetext
\begin{center}
\textbf{\large Supplementary Information}\\[6pt]
\blue{\textbf{\large Hyperfine spectroscopy and nuclear spin control of dark defects in diamond}}\\[6pt]
\blue{Alexander Ungar, Hao Tang, Andrew Stasiuk, Bo Xing, Boning Li, Ju Li, Alexandre Cooper, and Paola Cappellaro}
\end{center}
\setcounter{secnumdepth}{3}
\setcounter{section}{0} 
\setcounter{equation}{0}
\setcounter{figure}{0}
\setcounter{table}{0}
\setcounter{page}{1}
\makeatletter
\renewcommand{\thesection}{\arabic{section}}
\renewcommand{\theequation}{\arabic{equation}}
\renewcommand{\thefigure}{\arabic{figure}}
\renewcommand{\thetable}{\arabic{table}}
\renewcommand{\figurename}{Supplementary Figure}
\renewcommand{\tablename}{Supplementary Table}
\let\orig@seccntformat\@seccntformat
\renewcommand\@seccntformat[1]{\ifnum\pdfstrcmp{#1}{section}=0 Supplementary Note \thesection:\quad\else\orig@seccntformat{#1}\fi}
\crefalias{section}{suppnote}
\crefalias{figure}{suppfigure}
\crefalias{table}{supptable}
\crefalias{equation}{suppequation}
\begin{bibunit}[naturemag]
\section{Zero-field DEER signal}\label{sec:smZFDEER}
Here we derive the analytical form of the ZF-DEER signal measured in~\cref{fig:DEER}b,c to extract the hyperfine components of the X1 and X2 defects through the NV spin coherence signal. For this derivation we model the NV electronic spin as S = 1/2 and consider its coupling to a single electron-nuclear defect with S = 1/2 and I = 1/2 and uniaxial hyperfine interaction ($A_{xx} = A_{yy}= A_{\perp}$ and $A_{zz} = A_{\parallel}$). For this analysis, the NV is driven on resonance at the zero-field splitting $\Delta$, so we move to the rotating frame where the internal Hamiltonian $H_0^{\text{NV}}/(2\pi) = \Delta S_{z}^{\text{NV}} \rightarrow 0$. Under free evolution the dominant term in the NV Hamiltonian is the magnetic dipolar interaction with the X electronic spin, and we consider only its secular components (along $S_z^{\text{NV}}$). 

In the principal hyperfine frame of the X electron-nuclear system, its internal Hamiltonian is
\begin{align*}
    H_0^{\text{X}}/(2\pi) = A_{\perp}\left(S_x^{\text{X}}I_x^{\text{X}} + S_y^{\text{X}}I_y^{\text{X}}\right) + A_{\parallel}S_{z}^{\text{X}}I_{z}^{\text{X}}.
\end{align*}
Since at zero field the NV and X electronic spins are each aligned to its own defect axis, the secular dipolar Hamiltonian in the X principal frame becomes
\begin{align*}
    H_{\text{dip}}/(2\pi) = 2S_{z}^{\text{NV}}\left(d_{zx}S_{x}^{\text{X}} + d_{zy}S_{y}^{\text{X}} + d_{zz}S_{z}^{\text{X}}\right).
\end{align*}
For the time-dependent microwave drive we consider the coupling to all three electronic spin components and neglect the weaker nuclear spin coupling. The X microwave Hamiltonian is expressed as
\begin{align*}
    H_{\text{mw}}^{\text{X}}(t)/(2\pi) = 2(\Omega_xS_x + \Omega_yS_y + \Omega_zS_z)\cos(\omega_{\text{mw}} t).
\end{align*}

Under free evolution, $H_{\text{free}} = H_0^{\text{X}} + H_{\text{dip}}$, we treat the dipolar interaction as a perturbation since $|d| \ll |A|$. We move to the diagonal frame of $H_0^{\text{X}}$ expressed in the Bell basis defined by $\{\ket{\Phi_+},\ket{\Psi_+},\ket{\Psi_-},\ket{\Phi_-}\}$ with eigen-energies $ \{A_{\parallel}/4, (2A_{\perp}-A_{\parallel})/4, (-2A_{\perp}-A_{\parallel})/4, A_{\parallel}/4\}$. The three distinct transitions are
\begin{align*}
    \ket{\Psi_+} \rightleftharpoons \ket{\Phi_{\pm}} \; &\text{at} \;\omega_- = \frac{|A_{\parallel} - A_{\perp}|}{2}, \\
    \ket{\Psi_-} \rightleftharpoons \ket{\Phi_{\pm}} \; &\text{at} \;\omega_+ = \frac{|A_{\parallel} + A_{\perp}|}{2},\\
    \ket{\Psi_+} \rightleftharpoons \ket{\Psi_-} \; &\text{at} \;\omega_{\perp} = |A_{\perp}|.
\end{align*}
The above Hamiltonians $H_0^{\text{X}}$, $H_{\text{dip}}$, and $H_{\text{mw}}^{\text{X}}$ can then be transformed to the diagonal frame under the unitary transformation $U_b = e^{-i(\frac{\pi}{2})2S_y^{\text{X}}I_x^{\text{X}}}$, which leads to (omitting the factors of $2\pi$):
\begin{align*}
    H_0^b = & U_b^\dagger H_0U_b = A_{\parallel}S_zI_z + \frac{A_{\perp}}{2}\left(S_z-I_z\right), \\
    H_{\text{dip}}^b = & U_b^\dagger H_{\text{dip}}U_b = S_{z}^{\text{NV}}\left(4d_{zx}S_{z}I_{x} + 2d_{zy}S_{y} -4d_{zz}S_{x}I_{x}\right),\\
    H_{\text{mw}}^b(t) = & U_b^\dagger H_{\text{mw}}U_b = \left(4\Omega_{x}S_{z}I_{x} + 2\Omega_{y}S_{y} -4\Omega_{z}S_{x}I_{x}\right)\cos(\omega_{\text{mw}} t),
\end{align*}
where all $S$ and $I$ operators are now defined in the Bell basis, e.g. 
\begin{align*}
    S_z = 1/2\left(\ket{\Phi_+}\bra{\Phi_+} + \ket{\Psi_+}\bra{\Psi_+} - \ket{\Psi_-}\bra{\Psi_-} - \ket{\Phi_-}\bra{\Phi_-}\right)
\end{align*}

The DEER pulse sequence consists of a spin-echo on the NV along with a recoupling $\pi$-pulse on X, which is applied simultaneously with the NV $\pi$-pulse. The NV spin is first initialized to $\ket{0}$, while the X electronic and nuclear spins are both initially in fully mixed states, such that the initial density matrix is $\rho_0 = (\hat{I}_2 + \sigma_z)/2\otimes(\hat{I}_2/2)\otimes(\hat{I}_2/2) = (1/8)(\hat{I}_8 + 2S_z^{\text{NV}})$. The measured DEER signal is $S\propto \langle S_z^{\text{NV}}\rangle = \text{Tr}[S_z^\text{NV}U_{\text{DEER}}\rho_0U_{\text{DEER}}^{\dagger}]$. We account for the two NV $\pi/2$-pulses in the spin-echo by transforming the initial NV state to $\ket{+} = (\ket{0} + \ket{1})/\sqrt{2}$, and take the  measurement to be along the NV x-axis with final signal $S \propto \langle S_x^\text{NV}\rangle = \text{Tr}[S_x^\text{NV}U_{\text{DEER}}\rho_0U_{\text{DEER}}^{\dagger}]$. 

The total unitary for the sequence is then
\begin{align*}
    U_{\text{DEER}}= e^{-i(H_0^b - H_{\text{dip}}^b)\tau/2}U_{\pi}^\text{X}e^{-i(H_0^b +H_{\text{dip}}^b)\tau/2}, 
\end{align*}
where $U_{\pi}^\text{X}$ is the unitary for the $\pi$-pulse on the X electronic spin for the transition between two Bell states, as determined below. We consider the three cases for resonant microwave driving $\omega_{\text{mw}}$ at the transition frequencies $\{\omega_{\pm}, \omega_{\perp}\}$ and analyze the total DEER evolution in their respective rotating frames \blue{(written for $A_{\parallel} > A_{\perp} > 0$)}. 
\begin{enumerate}
    \item $\omega_{\text{mw}} = -\omega_{-}$:\\
    Here we analyze the dynamics in the subspace defined by $\{\ket{\Phi_+}, \ket{\Psi_+}\}$. To do so, we want to move to a frame where the transition frequency goes to zero and the microwave term coupling the two states becomes time independent under the Rotating Wave Approximation (RWA). For this, we move to the rotating frame defined by the unitary $U_{\omega_-}= e^{\red{-}i\omega_{-}t(S_z-I_z)}$. Dropping all time-dependent terms, the Hamiltonian under free evolution is
    \begin{align*}
        H_{
        \text{free},\pm}' \rightarrow & U_{\omega_-}^{\dagger}(H_0^b 
        \pm H_{\text{dip}}^b)U_{\omega_-} \red{+} \omega_{-}(S_z - I_z)\\
        &=\frac{A_{\parallel}}{2}(S_z - 2S_{\beta}I_z) \mp 2d_{zz}S_{z}^{\text{NV}}(S_{x}I_{x} - S_{y}I_{y}),
    \end{align*}
    where we define the electronic spin projection operators as $S_{\alpha,\beta} = (1/2)(\hat{I} \pm 2S_z)$. The transformed Hamiltonian under microwave drive for the $\pi$-pulse on X is
    \begin{align*}
        H_{\pi}' \rightarrow & U_{\omega_-}^{\dagger}(H_0^b 
        + H_{\text{mw}}^b(t))U_{\omega_-} \red{+} \omega_{-}(S_z - I_z)\\
        &=\frac{A_{\parallel}}{2}(S_z - 2S_{\beta}I_z) + 2\Omega_xS_zI_x + \Omega_yS_y\\
        \approx & \frac{A_{\parallel}}{2}(S_z - 2S_{\beta}I_z) + 2\Omega_xS_zI_x,
    \end{align*}
    where we neglect the dipolar terms since $|d| \ll |\Omega|$\red{. The static $\Omega_y$ term couples the off-resonant $\{\ket{\Phi_+}, \ket{\Psi_-}\}$ states, with detuning $|A_{\parallel}| \gg |\Omega|$, and the resonant $\{\ket{\Psi_+}, \ket{\Phi_-}\}$ states. Since $\ket{\Phi_\pm}$ are degenerate and the dipolar term couples them symmetrically, the $\Omega_y$ term only changes which superposition of $\ket{\Phi_\pm}$ is driven and the Rabi frequency, and does not change the signal derived below. We therefore set $\Omega_y = 0$}. The total unitary in the rotating frame is
    \begin{align*}
        U_{\text{DEER}}'&=e^{-iH_{\text{free},-}'(\tau/2)}e^{iH_{\pi}'t_{\pi}}e^{-iH_{
        \text{free},+}'(\tau/2)}\\
        &=U_{\text{free},-}^{'}(\tau/2)U_{\pi}^{'}U_{\text{free},+}^{'}(\tau/2).
    \end{align*}
     To flip the dipolar terms in $H_{\text{free},+}'$, the $\pi$-pulse duration should be equal to $t_{\pi} = \pi/\Omega_x$. We can drop the $H_0' = \frac{A_{\parallel}}{2}(S_z - 2S_{\beta}I_z)$ term from both $H_{\text{free},\pm}'$ and $H_{\pi}'$ since we assume the microwave term only affects dynamics in the $S_{\alpha}$ subspace (on resonance), and commutes with the dipolar terms in the $S_{\beta}$ subspace (off resonance). \red{We therefore} modify the final signal when tracing out the X electronic and nuclear spins to $S =(1/2 + \langle S_x^\text{NV}\rangle)/2$. The total unitary can then be simplified as
     \begin{align*}
         U_{\text{DEER}}' = & e^{-i(2d_{zz}S_{z}^{\text{NV}}(S_{x}I_{x} - S_{y}I_{y}))\tau/2}e^{-i(2\Omega_xS_zI_x)(\pi/\Omega_x)}e^{i(2d_{zz}S_{z}^{\text{NV}}(S_{x}I_{x} - S_{y}I_{y}))\tau/2}.
     \end{align*}
    Inserting the identity 
    \begin{align*}
        \hat{I} = (U_{\pi}^{'})^{\dagger}U_{\pi}^{'} = e^{i(2\Omega_xS_zI_x)(\pi/\Omega_x)}e^{-i(2\Omega_xS_zI_x)(\pi/\Omega_x)}
    \end{align*}
   on the right hand side of $U'_{\text{DEER}}$, the total unitary becomes
   \begin{align*}
       U_{\text{DEER}}'&=U_{\text{free},-}^{'}(\tau/2)U_{\pi}^{'}U_{\text{free},+}^{'}(\tau/2)(U_{\pi}^{'})^{\dagger}U_{\pi}^{'}.
   \end{align*}
   We can then simplify the $U_{\pi}^{'}U_{\text{free},+}^{'}(\tau/2)(U_{\pi}^{'})^{\dagger}$ term by transforming $H_{
        \text{free},+}'$ as follows:
   \begin{align*}
       U_{\pi}^{'}e^{i(2d_{zz}S_{z}^{\text{NV}}(S_{x}I_{x} -S_{y}I_{y}))\tau/2}(U_{\pi}^{'})^{\dagger} = e^{-i(2d_{zz}S_{z}^{\text{NV}}(S_{x}I_{x} +S_{y}I_{y}))\tau/2}.
   \end{align*}
   In the total unitary, the $S_{x}I_{x}$ terms add together and the $S_{y}I_{y}$ terms cancel, resulting in the final expression of
    \begin{align*}  
         U_{\text{DEER}}' &=e^{-i2d_{zz}S_{z}^{\text{NV}}S_{x}I_{x}\tau}e^{-i(2\Omega_xS_zI_x)(\pi/\Omega_x)}.
    \end{align*}
    The final signal can then be calculated as
    \begin{align*}
        S = & \frac{1}{2}\left(\frac{1}{2} + \langle S_x^\text{NV}\rangle \right),\\
        = & \frac{1}{2}\left(\frac{1}{2} + \text{Tr}[U_{\text{DEER}}'^{\dagger}S_x^\text{NV}U_{\text{DEER}}'\rho_0] \right), \\
        = & \frac{1}{2}\left(\frac{1}{2} + \text{Tr}[e^{(i2d_{zz}S_{z}^{\text{NV}}S_{x}I_{x}\tau)}S_x^\text{NV}e^{(-i2d_{zz}S_{z}^{\text{NV}}S_{x}I_{x}\tau)}\rho_0] \right).
    \end{align*}
    The operator $S_x^\text{NV}$ evolves under this unitary as 
    \begin{align*}
        S_x^\text{NV} \rightarrow \cos\left(\frac{d_{zz}\tau}{2}\right)S_x^\text{NV} +  4\sin\left(\frac{d_{zz}\tau}{2}\right)S_y^\text{NV}S_{x}I_{x}, 
    \end{align*}
    leading to the final signal of $S_-(\tau) = \frac{1}{4}\left(1 + \cos\left(\pi d_{zz}\tau\right)\right)$ (where we have reinserted the factor of $2\pi$ from the original dipolar Hamiltonian).

    \item $\omega_{\text{mw}} = \omega_{+}$:\\
    We follow the approach above, going to the rotating frame defined by the unitary $U_{\omega_+}= e^{i\omega_{+}t(S_z-I_z)}$. The Hamiltonian under free evolution becomes
    \begin{align*}
        H_{
        \text{free},\pm}' &=-\frac{A_{\parallel}}{2}(S_z - 2S_{\alpha}I_z) \mp 2d_{zz}S_{z}^{\text{NV}}(S_{x}I_{x} - S_{y}I_{y}), 
    \end{align*}
    achieving resonance in the $S_{\beta}$ manifold between the states $\{\ket{\Psi_-},\ket{\Phi_-}\}$. The time-independent microwave Hamiltonian during the $\pi$-pulse on X becomes
    \begin{align*}
        H_{\pi}' &=  -\frac{A_{\parallel}}{2}(S_z - 2S_{\alpha}I_z) + 2\Omega_xS_zI_x.
    \end{align*}
    After dropping the internal Hamiltonian term $H_0' = -\frac{A_{\parallel}}{2}(S_z - 2S_{\alpha}I_z)$ as above, the following analysis is identical to the $\omega_-$ case, and we recover the same signal $S_+(\tau) = \frac{1}{4}\left(1 + \cos\left(\pi d_{zz}\tau\right)\right)$.

    \item $\omega_{\text{mw}} = \omega_{\perp}$:\\
    We apply the rotating frame transformation under the unitary $U_{\omega_\perp}= e^{i\omega_{\perp}t(S_z-I_z)/2}$. The Hamiltonian under free evolution becomes
    \begin{align*}
        H_{
        \text{free},\pm}' &=A_{\parallel}S_zI_{z} \mp 2d_{zz}S_{z}^{\text{NV}}(S_{x}I_{x} - S_{y}I_{y}), 
    \end{align*}
    achieving resonance in the single-quantum manifold between the states $\{\ket{\Psi_+},\ket{\Psi_-}\}$. The time-independent microwave Hamiltonian becomes (keeping terms only in the single-quantum subspace)
    \begin{align*}
        H_{\pi}' &=A_{\parallel}S_zI_{z} - \Omega_z(S_{x}I_{x} + S_{y}I_{y}).
    \end{align*}
    Dropping the internal Hamiltonian term $H_0'$ as above, we see the microwave $\Omega_z(S_{x}I_{x} + S_{y}I_{y})$ term commutes with the dipolar $2d_{zz}(S_{x}I_{x} - S_{y}I_{y})$ term, so the dipolar evolution is canceled out in the total unitary $U_{\text{DEER}}'$. Therefore the NV spin state is unaffected, and the final signal under the DEER sequence is constant, with $S_{\perp}(\tau) = 1/2$.
\end{enumerate}
To summarize, the ZF-DEER signal for the two resonance cases when driving the \blue{$\ket{\Psi_{+}}\leftrightarrow \ket{\Phi_{\pm}}$ and $\ket{\Psi_{-}}\leftrightarrow \ket{\Phi_{\pm}}$ transitions at $\omega_{\text{mw}} = |A_{\parallel} - A_{\perp}|/2$ and $|A_{\parallel} + A_{\perp}|/2$, respectively,} is $S_{\pm}(\tau) = \frac{1}{4}\left(1 + \cos\left(\pi d_{zz}\tau\right)\right)$, and the ZF-DEER signal when driving the $\ket{\Psi_{+}}\leftrightarrow \ket{\Psi_{-}}$ transition at the resonance $\omega_{\text{mw}} = A_{\perp}$ is $S_{\perp}(\tau) = 1/2$ with zero contrast. Therefore sweeping the recoupling frequency of the X microwave drive when $\tau \sim 1/d_{zz}$ will lead to only two spectral dips at $|A_{\parallel} \pm A_{\perp}|/2$ with equal amplitude. \blue{Since these frequencies depend only on $|A_{\parallel} \pm A_{\perp}|$, the ZF-DEER spectrum is invariant under a change of the relative sign of $A_{\parallel}$ and $A_{\perp}$ and under their exchange. This ambiguity is resolved with NEETR (\cref{sec:smResid}).}

We have assumed above that the hyperfine interaction for the X defect is uniaxial. If instead $A_{xx} \ne A_{yy}$, then the six distinct zero-field transitions are
\begin{align*}
    \ket{\Psi_+} \rightleftharpoons \ket{\Phi_{+}} \; &\text{at} \;\omega_1 = \frac{|A_{zz} - A_{yy}|}{2}, \\
    \ket{\Psi_-} \rightleftharpoons \ket{\Phi_{+}} \; &\text{at} \;\omega_2 = \frac{|A_{zz} + A_{xx}|}{2},\\
    \ket{\Psi_+} \rightleftharpoons \ket{\Psi_-} \; &\text{at} \;\omega_{3} = \frac{|A_{xx} + A_{yy}|}{2},\\
    \ket{\Psi_+} \rightleftharpoons \ket{\Phi_{-}} \; &\text{at} \;\omega_4 = \frac{|A_{zz} - A_{xx}|}{2}, \\
    \ket{\Psi_-} \rightleftharpoons \ket{\Phi_{-}} \; &\text{at} \;\omega_5 = \frac{|A_{zz} + A_{yy}|}{2},\\
    \ket{\Phi_+} \rightleftharpoons \ket{\Phi_-} \; &\text{at} \;\omega_{6} = \frac{|A_{xx} - A_{yy}|}{2}.\\
\end{align*}
From the analysis above, the observable transitions with the DEER sequence would include $\omega_1, \omega_2, \omega_4$, and $\omega_5$ for a total of four spectral dips.

\section{Measurement uncertainty due to the geomagnetic field}\label{sec:smEarthB}
Here we estimate the frequency shifts due to Earth's magnetic field for the transitions in the zero-field DEER spectrum~(\cref{fig:DEER}b). At zero field the resonance frequencies are
\begin{align*}
    \omega_{\mathrm{e}}^0/(2\pi)
    &= \frac{|A_{\parallel}\pm A_{\perp}|}{2}.
\end{align*}
The estimated magnitude of the geomagnetic field at the position of the sample is $B_{\mathrm{e}} \approx 0.5$ G. Since the relative alignment between the field and the defect axis is unknown, the upper bound on the frequency shift for each transition can be estimated by considering the Hamiltonian when the field is perfectly aligned or anti-aligned:
\begin{align*}
     H_{\pm}/(2\pi) = A_{\perp}\left(S_xI_x + S_yI_y\right) + A_{\parallel}S_{z}I_{z} \pm \gamma_{\mathrm{e}}B_{\mathrm{e}}S_z.
\end{align*}
The resonance frequencies with the aligned or anti-aligned external field become
\begin{align*}
    \omega_{\mathrm{e}}(B_{\mathrm{e}+})/({2\pi})
    &= \frac{1}{2}\left(A_{\parallel} \pm \gamma_{\mathrm{e}}B_{\mathrm{e}} \pm \sqrt{A_{\perp}^2 + \left(\gamma_{\mathrm{e}}B_{\mathrm{e}}\right)^2}\right),\\
    \omega_{\mathrm{e}}(B_{\mathrm{e}-})/(2\pi)
    &= \frac{1}{2}\left(A_{\parallel} \mp \gamma_{\mathrm{e}}B_{\mathrm{e}} \pm \sqrt{A_{\perp}^2 + \left(\gamma_{\mathrm{e}}B_{\mathrm{e}}\right)^2}\right).
\end{align*}
\red{In both cases, each zero-field transition splits into two components separated by $\gamma_{\mathrm{e}}B_{\mathrm{e}}$, while the change of the square-root term is of second order in $\gamma_{\mathrm{e}}B_{\mathrm{e}}/A_{\perp}$,} leading to an uncertainty due to the geomagnetic field of $\pm \gamma_{\mathrm{e}}B_{\mathrm{e}}/2 \approx \pm 0.7 \, \text{MHz}$. 

The total frequency uncertainty includes both the geomagnetic shift and the spectral linewidth. We estimate the spectral linewidth of each transition in the ZF-DEER spectrum using the half width at half maximum (HWHM) where $\gamma \approx 0.2 \, \text{MHz}$. The total uncertainty for each resonance is then $\pm\sqrt{0.7^2 + 0.2^2}$, leading to an uncertainty in the hyperfine components  for $A_{\parallel}$ and $A_{\perp}$ of $\Delta A \approx 1 \, \text{MHz}$.

\section{Zero-field DEER measurements on additional defects}\label{sec:smMoredefects}
\begin{figure}[ht]
	\centering
	\includegraphics[width=6.8in]{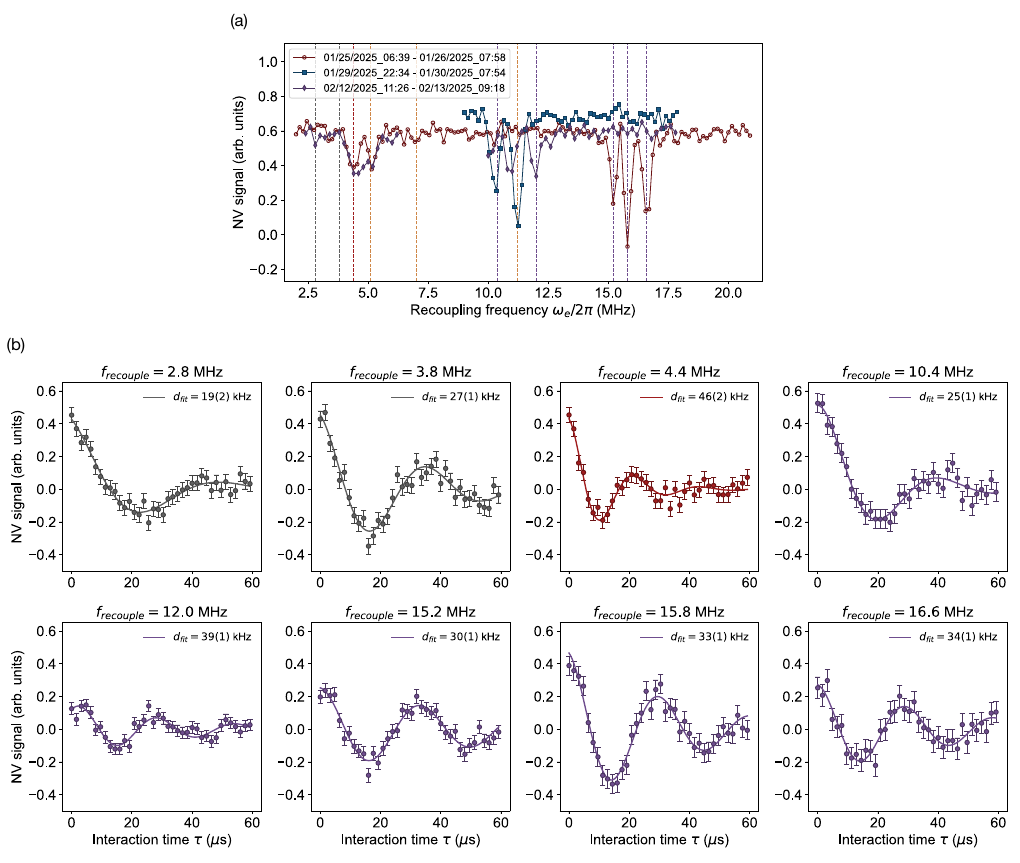}
	\caption{ZF-DEER characterization of other resonances (distinct from the X1 and X2 frequencies) indicating additional defects in the NV environment. (a) Measurement of the ZF-DEER spectrum~(\cref{fig:DEER}b) at $\tau= 12 \,\mu\text{s}$ over three different averaging windows reveals several peaks with dynamic behavior. For each trace, we average for a minimum of approximately 10 hours to achieve sufficient SNR. The scans in blue and purple investigate the shifting resonances between 10-17 MHz, possibly due to a defect with a fluctuating charge state. The dashed lines indicate the resonances found in the main-text scan or from the additional scans shown here. Orange designates the X1 and X2 transitions. For the additional transitions, gray designates the peaks found only in the main-text scan, red for the static peak, and purple for the dynamic peaks. (b) We scan the interaction time in the ZF-DEER sequence with recoupling frequencies set to the resonances at the dashed lines (excluding the X1 and X2 transitions) to distinguish different defects based on their coupling strengths. Coupling strengths $d$ are estimated from fitting to an exponentially decaying cosine function. The traces at 15.2, 15.8, and 16.6 MHz have consistent coupling and appear within the same frequency scan in (a), suggesting they belong to the same defect.}
	\label{fig:supp_ZFSEDORother}
\end{figure}
In the ZF-DEER spectrum of~\cref{fig:DEER}b we observe additional resonances which are not associated with the X1 and X2 defect transitions. Our attempts  to identify the origin of these resonances and assign them to additional defects coupled to the NV have been complicated by the fact that they appear to fluctuate over different days of measurements. To better understand their origin and dynamics, we perform additional ZF-DEER scans over the same frequency range probed in the main text (see~\cref{fig:supp_ZFSEDORother}a). Over three independent averaging windows taken across different days we find several new peaks (at 15.2, 15.8, and 16.6 MHz) which are not present in the main-text scan. Additionally, we find the two resonances at 2.8 and 3.8 MHz present in the main-text scan are missing in the additional scans. Several of the transitions above 10 MHz appear to dynamically disappear and reappear across the three averaging windows. Furthermore, some of these dynamical peaks appear to be anti-correlated with each other. The group of peaks at 10.4, 11.2, and 12.0 MHz appear to jump to the peaks at 15.2, 15.8, and 16.6 MHz between the different scans. This  possibly suggests that one defect is hopping between multiple charge or Jahn-Teller states~\cite{Degen2021}. In a previous work, we detected an additional defect labeled as X3 using DEER, which has a hyperfine splitting of approximately 3 MHz~\cite{Ungar2024}. In these prior measurements, it was also noticed that the resonances would disappear and reappear randomly over the timescale of several days. This strongly suggests that a subset of these resonances observed at zero field belongs to the spectrally unstable X3 defect.

To assign these peaks to additional distinct defects, we measure the electronic spin coupling to the NV at each identified resonance, as in~\cref{fig:DEER}c. For each additional peak found in both the main-text spectrum of~\cref{fig:DEER}b and in the scans from~\cref{fig:supp_ZFSEDORother}a, we drive on resonance and sweep the interaction time, observing coherent oscillations of the NV spin state~(\cref{fig:supp_ZFSEDORother}b). The signals indicate that these peaks correspond to the resonant driving of single electron-nuclear spin transitions. The electronic spin coupling strengths can be estimated from the signal's oscillation frequency. Following the approach of the main text, we search for pairs of signals with matching coupling strengths to assign each resonance to a specific defect. We find the resonances at 3.8 and 10.4 MHz have matching coupling strengths, but it is inconclusive whether they belong to the same defect as the peaks do not simultaneously appear within a given averaging window. Further, the set of three peaks at 15.2, 15.8, and 16.6 MHz (which appear together in the first scan) have matching coupling strengths, indicating they could belong to the same defect. The presence of a third resonance suggests the defect may have a non-uniaxial hyperfine interaction.

We further verify that none of the resonances in the new scans of~\cref{fig:supp_ZFSEDORother}b correspond to additional transitions of the X1 or X2 defects. We first note that the time trace for the 4.4 MHz signal features an oscillation frequency of $46\, \text{kHz}$, matching the X2 coupling strength as measured in the main text. While this could possibly represent an additional X2 transition resulting from a \red{non-uniaxial} hyperfine interaction (where $A_{xx}\ne A_{yy}$), we note that the coherence time of this transition is $13(2)\,\mu\text{s}$, which is not consistent with the X2 transition coherence times of 30(7) and 36(8) $\mu$s. This indicates that this additional transition does not belong to the X2 defect. The remaining resonances investigated in \cref{fig:supp_ZFSEDORother}b do not feature oscillation frequencies which are consistent with the X1 or X2 coupling strengths. Therefore, we conclude that both X1 and X2 have a maximum of two distinguishable transitions at zero field \blue{within our scan range}, suggesting both defects are approximately uniaxial.

\section{Angular dependence of the hyperfine interaction \blue{and nuclear spin transition frequencies}}\label{sec:smEigen}
For an electron-nuclear defect with uniaxial symmetry ($A_{xx} = A_{yy}$), the hyperfine tensor in its principal frame is represented by the diagonal matrix
\begin{align*}
    \hat{\mathbf{A}}_{\mathrm{princ}} = \text{diag}[A_{\perp}, A_{\perp}, A_{\parallel}].
\end{align*}
The axis defining the principal frame (along $A_{\parallel}$) is parameterized by two angles relative to the crystal coordinate frame $\{\theta_{\text{X}}, \phi_{\text{X}}\}$ where $\theta_{\text{X}}$
is the polar angle from [001] and $\phi_{\text{X}}$ is the azimuthal angle from [100] towards [010].  To transform the hyperfine matrix into another coordinate frame, we use the following rotation matrix defined by the Euler angles $\{\alpha, \beta, \gamma\ = 0\}$~\cite{Degen2021}:
\begin{align*}
    \hat{R}(\alpha, \beta) = \begin{pmatrix}
                \cos(\beta)\cos(\alpha) &  \cos(\beta)\sin(\alpha) & -\sin(\beta)\\
                -\sin(\alpha) &  \cos(\alpha) & 0\\
                \sin(\beta)\cos(\alpha) &  \sin(\beta)\sin(\alpha) & \cos(\beta)
            \end{pmatrix}.
\end{align*}
The DEER measurements to detect the X1 and X2 defects are performed with the external field aligned along the NV [111] axis~(\cref{fig:DEER}a).  To transform $\hat{\mathbf{A}}_{\mathrm{princ}}$ into this frame, we apply the following two rotations:
\begin{align*}
    \hat{\mathbf{A}}' = R(\phi_{\text{NV}},\theta_{\text{NV}})R^{T}(\phi_{\text{X}},\theta_{\text{X}})\hat{\mathbf{A}}_{\mathrm{princ}} R(\phi_{\text{X}},\theta_{\text{X}})R^{T}(\phi_{\text{NV}},\theta_{\text{NV}}),
\end{align*}
where $\theta_{\text{NV}} = 54.7^{\circ}$ and $\phi_{\text{NV}} = 45^{\circ}$. The full hyperfine Hamiltonian for the X defect in this frame is $\mathbf{S}\cdot\hat{\mathbf{A}}'\cdot\mathbf{I}$. Since $|\gamma_{\mathrm{e}}B_0| \gg |A_{\parallel}|, |A_{\perp}|$ we keep only the secular (energy conserving) terms, leading to the Hamiltonian of \cref{eq:H0_B} in the main text. The secular hyperfine components are:
\begin{align}\label{eq:Azi}
\begin{split}
A_{zx} &= \hat{\mathbf{e}}_k\cdot\hat{\mathbf{A}}'\cdot \hat{\mathbf{e}}_i \\
&= -\tfrac{1}{8} (A_{\parallel}-A_{\perp}) \Big( 
      -4 \sin (2 \theta_{\text{X}}) \cos (2 \theta_{\text{NV}}) 
        \cos (\phi_{\text{NV}}-\phi_{\text{X}}) \\ 
        &\qquad\qquad +\sin (2 \theta_{\text{NV}}) \big(
          \cos (2 \theta_{\text{X}} ) 
          (\cos (2 (\phi_{\text{NV}}- \phi_{\text{X}} ))+3)
          +2 \sin ^2(\phi_{\text{NV}}-\phi_{\text{X}} )\big)\Big),\\
A_{zy} &= \hat{\mathbf{e}}_k\cdot\hat{\mathbf{A}}'\cdot \hat{\mathbf{e}}_j \\
&= - (A_{\parallel} - A_{\perp}) 
   \sin(\theta_{\text{X}}) \sin(\phi_{\text{NV}}-\phi_{\text{X}}) \\
&\quad \times \Big(
      \sin(\theta_{\text{NV}}) \sin(\theta_{\text{X}}) \cos(\phi_{\text{NV}}-\phi_{\text{X}})
      + \cos(\theta_{\text{NV}}) \cos(\theta_{\text{X}})
   \Big),\\
A_{zz} &= \hat{\mathbf{e}}_k\cdot\hat{\mathbf{A}}'\cdot \hat{\mathbf{e}}_k \\
   &= \cos^2(\theta_{\text{NV}})
   \big( A_{\parallel} \cos^2(\theta_{\text{X}}) 
       + A_{\perp} \sin^2(\theta_{\text{X}}) \big) \\[6pt]
&\quad + \tfrac{1}{4} \sin^2(\theta_{\text{NV}}) 
   \Big( A_{\parallel} + 3A_{\perp} 
       - (A_{\parallel} - A_{\perp}) 
         \big( \cos(2\theta_{\text{X}})
             - 2 \cos\!\big(2(\phi_{\text{NV}}-\phi_{\text{X}})\big) 
               \sin^2(\theta_{\text{X}}) \big) \Big) \\[6pt]
&\quad + (A_{\parallel} - A_{\perp}) 
   \cos(\theta_{\text{NV}})\cos(\phi_{\text{NV}}-\phi_{\text{X}})
   \sin(\theta_{\text{NV}})\sin(2\theta_{\text{X}}).
\end{split}
\end{align}
\blue{The components $A_{zi}$ above hold for either sign of $A_{\parallel}$ and $A_{\perp}$. Equivalently, defining the unit vector $\hat{\mathbf{n}}$ along the principal hyperfine axis and the angle $\vartheta$ between $\hat{\mathbf{n}}$ and the field direction $\hat{\mathbf{z}}$ ($\cos\vartheta = \hat{\mathbf{n}}\cdot\hat{\mathbf{z}}$), the secular hyperfine vector, given by the third row of $\hat{\mathbf{A}}'$, is
\begin{align}\label{eq:avec}
\mathbf{a}_z \equiv (A_{zx}, A_{zy}, A_{zz}) = A_{\perp}\hat{\mathbf{z}} + (A_{\parallel}-A_{\perp})\cos\vartheta\,\hat{\mathbf{n}},
\end{align}
with
\begin{align}\label{eq:A}
A^2 \equiv |\mathbf{a}_z|^2 = A_{\parallel}^2\cos^2\vartheta + A_{\perp}^2\sin^2\vartheta, \qquad A_{zz} = A_{\parallel}\cos^2\vartheta + A_{\perp}\sin^2\vartheta,
\end{align}
where $A$ is the hyperfine splitting measured with DEER (\cref{fig:DEER}a).}

\subsection{\blue{Nuclear spin transition frequencies}}
\blue{Within the secular approximation of \cref{eq:H0_B} in the main text, the nuclear spin in the electronic manifold $m_{\mathrm{s}} = \pm 1/2$ experiences the effective field $\mathbf{h}_{\pm} = \pm\tfrac{1}{2}\mathbf{a}_z - \gamma_{\mathrm{n}}B_0\hat{\mathbf{z}}$, giving the nuclear spin transition frequencies $f_{\pm} = |\mathbf{h}_{\pm}|$:
\begin{align}\label{eq:nucfreq}
f_{\pm} = \frac{1}{2}\sqrt{A^2 \mp 4\gamma_{\mathrm{n}}B_0A_{zz} + 4(\gamma_{\mathrm{n}}B_0)^2} \approx \frac{A}{2} \mp \frac{A_{zz}}{A}\gamma_{\mathrm{n}}B_0 + \frac{(\gamma_{\mathrm{n}}B_0)^2}{A}\left(1 - \frac{A_{zz}^2}{A^2}\right)
\end{align}
to second order in $\gamma_{\mathrm{n}}B_0/A$. The NEETR resonances $\omega_{\mathrm{n}\pm}/(2\pi)$ of the main text correspond to the larger and smaller of $f_{\pm}$. The second-order term is the same in both electronic manifolds and cancels in the nuclear frequency splitting,
\begin{align}\label{eq:dwn1}
\Delta\omega_{\mathrm{n}}/(2\pi) = |f_- - f_+| \approx \frac{2|\gamma_{\mathrm{n}}B_0A_{zz}|}{A},
\end{align}
which therefore holds up to corrections of third order in $\gamma_{\mathrm{n}}B_0/A$. In the regime $A > 2|\gamma_{\mathrm{n}}B_0|$ relevant here, the two allowed electronic transitions measured with DEER are separated by $f_+ + f_-$, which equals the hyperfine splitting $A$ up to corrections of order $(\gamma_{\mathrm{n}}B_0)^2/A$.}

\subsection{\blue{Dependence on the relative sign of the hyperfine components}}
\blue{\cref{eq:A} shows that the hyperfine splitting $A$ measured with DEER depends only on $A_\parallel^2$ and $A_\perp^2$. Therefore, a change of the relative sign of $A_\parallel$ and $A_\perp$ leaves unchanged both the observed $A$ at any field orientation  and the ZF-DEER transition frequencies $|A_\parallel\pm A_\perp|/2$ (\cref{sec:smZFDEER}). Exchanging $A_\parallel$ and $A_\perp$ also leaves the ZF-DEER frequencies unchanged. Given the magnitudes $|A_\parallel|$ and $|A_\perp|$ from ZF-DEER, and the measured hyperfine splitting $A$ from DEER we can determine the orientation through
\begin{align}\label{eq:theta}
\cos^2\vartheta = \frac{A^2 - A_\perp^2}{A_\parallel^2 - A_\perp^2}
\end{align}
for each of the four possible assignments, corresponding to the two relative signs and to either magnitude lying along the principal axis. In contrast, $A_{zz}$ depends on the relative sign. For same-sign components, $A_{zz}/A$ is bounded from below for all orientations,
\begin{align}
\frac{A_{zz}}{A} \geq \frac{2\sqrt{A_\parallel A_\perp}}{|A_\parallel|+|A_\perp|} \qquad (A_\parallel A_\perp > 0),
\end{align}
which evaluates to 0.976 for X1 and 0.891 for X2, such that $\Delta\omega_{\mathrm{n}}/(2\pi)\approx 2|\gamma_{\mathrm{n}}|B_0$ is an accurate approximation. For opposite-sign components, $A_{zz}$ vanishes at $\tan^2\vartheta = -A_\parallel/A_\perp$, and $|A_{zz}|/A$ can take any value between zero and one depending on the orientation. In this case, $\Delta\omega_{\mathrm{n}}$ can be substantially reduced compared to $2|\gamma_{\mathrm{n}}|B_0$, and the measured splitting carries information on the orientation of the principal axis as well as the relative sign of the hyperfine components.}

\subsection{\blue{Correction from the non-secular hyperfine terms}}
\blue{The non-secular hyperfine terms $S_x(\mathbf{a}_x\cdot\mathbf{I}) + S_y(\mathbf{a}_y\cdot\mathbf{I})$, where the non-secular hyperfine vectors $\mathbf{a}_x$ and $\mathbf{a}_y$ are the first two rows of $\hat{\mathbf{A}}'$, couple the two electronic manifolds and shift the nuclear spin transition frequencies. Second-order perturbation treatments of the spin Hamiltonian for anisotropic interactions and arbitrary orientations are given in Refs.~\cite{Rockenbauer1973, Iwasaki1974}. For the nuclear frequency splitting, the effect of these terms follows more directly from an exact property of the full Hamiltonian $H/(2\pi) = \gamma_{\mathrm{e}}B_0S_z + \mathbf{S}\cdot\hat{\mathbf{A}}'\cdot\mathbf{I} - \gamma_{\mathrm{n}}B_0I_z$, without perturbation theory. Its eigenvalues are $\epsilon\pm\tfrac{1}{2}f_+$ in the upper electronic manifold ($m_{\mathrm{s}} = +1/2$) and $-\epsilon\pm\tfrac{1}{2}f_-$ in the lower one, where $f_\pm$ are the exact nuclear spin transition frequencies and the two manifolds are centered at $\pm\epsilon$ because $H$ is traceless. The sum of the cubed eigenvalues is then
\begin{align}
\mathrm{tr}\left[(H/2\pi)^3\right] = \tfrac{3}{2}\,\epsilon\left(f_+^2 - f_-^2\right).
\end{align}
The same trace can be evaluated directly by writing $\mathbf{S} = \boldsymbol{\sigma}/2$ and $\mathbf{I} = \boldsymbol{\tau}/2$ with the Pauli matrices $\boldsymbol{\sigma}$ and $\boldsymbol{\tau}$. Only two types of terms have a nonzero trace: the products of the electron Zeeman, nuclear Zeeman, and secular hyperfine terms, and the terms cubic in the hyperfine interaction, which with $\mathrm{tr}(\sigma_i\sigma_j\sigma_k) = 2i\varepsilon_{ijk}$ are proportional to $\det(\hat{\mathbf{A}}')$. This gives
\begin{align}
\mathrm{tr}\left[(H/2\pi)^3\right] = -\tfrac{3}{2}\left[\gamma_{\mathrm{e}}B_0\,\gamma_{\mathrm{n}}B_0\,A_{zz} + \tfrac{1}{4}\det(\hat{\mathbf{A}})\right],
\end{align}
where $\det(\hat{\mathbf{A}}') = \det(\hat{\mathbf{A}}) = A_\perp^2A_\parallel$ is invariant under rotation and therefore independent of the defect orientation. Combining the two expressions yields the exact relation
\begin{align}\label{eq:trace}
f_+^2 - f_-^2 = -\frac{\gamma_{\mathrm{e}}B_0\,\gamma_{\mathrm{n}}B_0\,A_{zz} + \det(\hat{\mathbf{A}})/4}{\epsilon}.
\end{align}
For $\epsilon = \gamma_{\mathrm{e}}B_0/2$, the first term gives the secular splitting of \cref{eq:dwn1}, and the second term is the correction from the non-secular terms. Since $\epsilon = \gamma_{\mathrm{e}}B_0/2$ up to a relative correction of order $(A/\gamma_{\mathrm{e}}B_0)^2$ and $f_+ + f_- \approx A$, the nuclear frequency splitting $\Delta\omega_{\mathrm{n}}/(2\pi) = |f_+^2 - f_-^2|/(f_+ + f_-)$ becomes
\begin{align}\label{eq:dwn_sm}
\Delta\omega_{\mathrm{n}}/(2\pi) \approx 2\left|\gamma_{\mathrm{n}}B_0\frac{A_{zz}}{A} + \frac{\det(\hat{\mathbf{A}})}{4\gamma_{\mathrm{e}}B_0A}\right|, \qquad \det(\hat{\mathbf{A}}) = A_\perp^2A_\parallel,
\end{align}
which is \cref{eq:dwn} of the main text. Over all orientations, \cref{eq:dwn_sm} agrees with exact diagonalization to within 15~kHz for the fitted hyperfine tensors of X1 and X2. For an isotropic hyperfine interaction ($A_\parallel = A_\perp$), \cref{eq:dwn_sm} reduces to the known second-order result $\Delta\omega_{\mathrm{n}}/(2\pi) = 2|\gamma_{\mathrm{n}}B_0 + A^2/(4\gamma_{\mathrm{e}}B_0)|$, obtained by expanding the exact Breit--Rabi solution~\cite{Breit1931, Schweiger2001}, in which the correction enlarges the splitting for $\gamma_{\mathrm{n}}>0$ and reduces it for $\gamma_{\mathrm{n}}<0$. The two terms add when $\gamma_{\mathrm{n}}A_{zz}$ and $A_\parallel$ have the same sign, and partially cancel otherwise. A change of the overall sign of the hyperfine tensor reverses both terms and leaves $\Delta\omega_{\mathrm{n}}$ unchanged, so that only the relative sign of the hyperfine components can be determined.}

\blue{As an example, consider X1 with a $^1$H nuclear spin, for which $\gamma_{\mathrm{e}}B_0 = 1020.1$~MHz and $\gamma_{\mathrm{n}}B_0 = 1.550$~MHz at $B_0 = 364$~G. For the best-fit solution (i), $(A_\parallel, A_\perp) = (+38.7, -25.8)$~MHz at $\vartheta = 78.0^\circ$ (\cref{tab:signfit}), \cref{eq:A} gives $A = 26.47$~MHz and $A_{zz} = -23.01$~MHz, so that the two terms in \cref{eq:dwn_sm} are $-1.35$~MHz and $+0.24$~MHz. \cref{eq:dwn_sm} then yields $\Delta\omega_{\mathrm{n}}/(2\pi) = 2.22$~MHz, in agreement with the value of 2.21~MHz from exact diagonalization and with the measured splitting of 2.217~MHz. In contrast, for same-sign components $(+38.7, +24.8)$~MHz, with $\vartheta$ fixed by the measured $A = f_+ + f_- = 26.51$~MHz through \cref{eq:theta}, \cref{eq:dwn_sm} yields 3.50~MHz.}

\section{\blue{Determination of the relative hyperfine sign and nuclear spin species}}\label{sec:smResid}
\blue{We determine the signed hyperfine components of each defect with a global fit of all measured transition frequencies: the two ZF-DEER frequencies (\cref{fig:DEER}b), the hyperfine splitting $A$ from DEER (\cref{fig:DEER}a), and the two NEETR frequencies $\omega_{\mathrm{n}\pm}$ (\cref{fig:NVXENDOR}b,c). Model frequencies are computed by exact diagonalization of the full spin Hamiltonian (\cref{sec:smEigen}) at $B_0 = 364$~G, assuming a uniaxial hyperfine tensor and an isotropic $g$-tensor, with free parameters $\{A_\parallel, A_\perp, \vartheta\}$. We minimize
\begin{align}\label{eq:chi2}
\chi^2 = \sum_k \left(\frac{\nu_k^{m} - \nu_k(A_\parallel, A_\perp, \vartheta)}{\sigma_k}\right)^2,
\end{align}
where $\nu_k^m$ are the measured frequencies with uncertainties $\sigma_k$. For the ZF-DEER frequencies, these combine the geomagnetic shift of 0.7~MHz and the HWHM of each resonance (0.2~MHz) in quadrature, giving 0.73~MHz (\cref{sec:smEarthB}). For $A$, they combine the HWHM of the two DEER resonances in quadrature (0.5~MHz for X1 and 0.4~MHz for X2)\recent{. For the NEETR frequencies, they are} the HWHM of the resonances (30--40~kHz for X1 and 8--10~kHz for X2). The X2 DEER spectrum was recorded at a field higher by $\approx 2$~G, which changes the hyperfine splitting $A$ by less than 1~kHz, and all NEETR spectra were recorded at 364~G. With five measured frequencies and three free parameters, each fit has two degrees of freedom. We repeat the fit for the four possible assignments of the ZF-DEER magnitudes: same or opposite relative sign, with either the larger or the smaller magnitude along the principal axis. The results are listed in~\cref{tab:signfit}.}

\begin{table}[ht]
\centering
\blue{\begin{tabular}{llcccc}
\toprule
Defect (nucleus) & Hypothesis & $A_\parallel$ (MHz) & $A_\perp$ (MHz) & $\vartheta$ & $\chi^2$\\\midrule
X1 ($^1$H) & same sign, $|A_\parallel|>|A_\perp|$ & $+37.1$ & $+22.3$ & $62^\circ$ & 529\\
 & same sign, $|A_\parallel|<|A_\perp|$ & $+22.3$ & $+36.2$ & $29^\circ$ & 740\\
 & opposite sign, $|A_\parallel|>|A_\perp|$ & $+38.7$ & $-25.8$ & $78^\circ$ & 1.0\\
 & opposite sign, $|A_\parallel|<|A_\perp|$ & $+23.4$ & $-38.4$ & $24^\circ$ & 2.0\\\midrule
X1 ($^{19}$F) & same sign, $|A_\parallel|>|A_\perp|$ & $+37.4$ & $+22.3$ & $63^\circ$ & 378\\
 & same sign, $|A_\parallel|<|A_\perp|$ & $+22.3$ & $+36.2$ & $29^\circ$ & 559\\
 & opposite sign, $|A_\parallel|>|A_\perp|$ & $+38.7$ & $-26.1$ & $81^\circ$ & 1.7\\
 & opposite sign, $|A_\parallel|<|A_\perp|$ & $+23.6$ & $-38.4$ & $23^\circ$ & 1.4\\\midrule
X2 ($^{15}$N) & same sign, $|A_\parallel|>|A_\perp|$ & $+15.8$ & $+6.3$ & $65^\circ$ & 2.0\\
 & same sign, $|A_\parallel|<|A_\perp|$ & $+6.3$ & $+13.8$ & $30^\circ$ & 17\\
 & opposite sign, $|A_\parallel|>|A_\perp|$ & $+16.5$ & $-7.8$ & $74^\circ$ & 2.9\\
 & opposite sign, $|A_\parallel|<|A_\perp|$ & $+8.6$ & $-14.1$ & $9^\circ$ & 47\\
\bottomrule
\end{tabular}}
\caption{\blue{Global fits of the ZF-DEER, DEER, and NEETR frequencies for the four assignments of the relative sign and ordering of the hyperfine components. Best-fit parameters are listed for each hypothesis, with $\vartheta$ the angle between the principal hyperfine axis and the external field along the NV [111] axis. The overall sign of each tensor is undetermined. For the same-sign hypotheses of X1, the minimum lies at the edge of the search range ($\pm2.5$~MHz around the ZF-DEER magnitudes), and $\chi^2$ remains large throughout. Each fit has two degrees of freedom.}}
\label{tab:signfit}
\end{table}

\blue{For X1, the same-sign hypotheses are excluded ($\chi^2 > 350$ for both $^1$H and $^{19}$F): since $A_{zz}/A \geq 0.976$ for all orientations (\cref{sec:smEigen}), the predicted nuclear splitting of 3.5--3.7~MHz for $^1$H significantly exceeds the measured 2.217~MHz. Both opposite-sign hypotheses reproduce all five frequencies within uncertainty ($\chi^2 < 3$) and cannot be distinguished with the present measurements: solution (i) gives a slightly lower $\chi^2$, but the difference ($\Delta\chi^2 = 1.0$) is not significant. The two solutions correspond to isotropic and dipolar contributions $(a_{\text{iso}}, T) = (-4.3, 21.5)$~MHz and $(-17.8, 20.6)$~MHz, respectively, with $a_{\text{iso}} = (A_\parallel + 2A_\perp)/3$ and $T = (A_\parallel - A_\perp)/3$, up to an overall sign.}

\blue{The fit also constrains the orientation of the principal axis. Repeating the fit with $\vartheta$ fixed, we find $\Delta\chi^2 < 4$ for $\vartheta\approx 76.5$--$79.5^\circ$ for solution (i) and $\vartheta\approx 23$--$24^\circ$ for solution (ii). These ranges exclude a principal axis along any $\langle 111\rangle$, $\langle 110\rangle$, $\langle 211\rangle$, or $\langle 100\rangle$ direction, whereas a $\langle 221\rangle$ direction ($\vartheta = 78.9^\circ$), which lies in a $\{110\}$ mirror plane, is allowed. X1 therefore has a low-symmetry structure (C$_{1h}$ or C$_1$), and trigonal centers with a $\langle 111\rangle$ axis, including those with motionally averaged trigonal symmetry, are excluded. This analysis assumes the external field is aligned along [111]. A misalignment of a few degrees would broaden the allowed ranges accordingly.}

\blue{The global fit further assumes a uniaxial hyperfine tensor, as indicated by the two ZF-DEER transitions of X1. A non-uniaxial tensor ($A_{xx} \neq A_{yy}$) would split each transition into a doublet separated by $|A_{xx} - A_{yy}|/2$ (\cref{sec:smZFDEER}). The other resonances in the scanned range of 1--35~MHz have NV coupling strengths of at most 46~kHz, distinct from the 70~kHz of X1 (\cref{sec:smMoredefects}), so that no additional X1 transition is observed in this range. Together with the width of the ZF-DEER transitions (HWHM 0.2~MHz), this limits the deviation from a uniaxial tensor to $|A_{xx} - A_{yy}| \lesssim 1$~MHz\recent{. Larger deviations are possible only if both additional transitions fall outside the scanned range, which requires a difference of $\approx 12$--16~MHz between two principal values or a principal value larger than $\approx 45$~MHz. The present measurements cannot exclude this.} For the calculated structures, however, the predicted transitions that fall within the range of the scan are inconsistent with the observed spectrum (\cref{sec:smDFT}).}

\blue{To identify the nuclear spin species without assuming a specific isotope, we repeat the analysis with $\gamma_{\mathrm{n}}$ as a free parameter using \cref{eq:dwn_sm}, allowing for the uncertainties of the hyperfine magnitudes and of the measured splitting. For the same-sign hypotheses, the measured splitting would require $\gamma_{\mathrm{n}}\approx 1.9$--2.7~kHz/G or $\gamma_{\mathrm{n}}\approx -3.6$ to $-4.2$~kHz/G. No light $\text{I}=1/2$ isotope lies in these ranges, and the only candidates, the thallium isotopes ($\gamma_{\mathrm{n}}\approx 2.4$~kHz/G), are not plausible impurities. For the opposite-sign hypotheses, the measurements are consistent with $^1$H, $^{19}$F, $^3$H, and $^3$He (and, for solution (ii), also with thallium), whereas $^{13}$C, $^{15}$N, $^{29}$Si, and $^{31}$P are excluded. Of these, only $^1$H and $^{19}$F are plausible impurities in CVD diamond, and the two give comparable $\chi^2$~(\cref{tab:signfit}). We assign X1 as hydrogen-related given the much higher concentration of hydrogen incorporated during CVD growth (see main text).}

\blue{For X2, the same-sign hypothesis with $|A_\parallel| > |A_\perp|$ provides the best fit ($\chi^2 = 2.0$), consistent with a $^{15}$N nuclear spin and with the same-sign components calculated for the WAR9 structure (\cref{sec:smDFT}). The opposite-sign hypothesis with $|A_\parallel| > |A_\perp|$ ($\chi^2 = 2.9$) is only slightly disfavored and cannot be excluded. Unlike for X1, the opposite-sign hypotheses for X2 allow $A_{zz}$ to approach zero within the experimental uncertainty, so that NEETR alone does not constrain the nuclear species in this case. The assignment of X2 as $^{15}$N-related is instead supported by the same-sign fit, the $^{15}$N implantation of the sample, and the agreement with the calculated hyperfine components of WAR9.}

\blue{More generally, a single NEETR spectrum identifies the nuclear spin species only up to a discrete set of candidates, one for each assignment of the relative sign and ordering of the hyperfine components. The remaining ambiguity, including the choice between the two opposite-sign solutions for X1, \recent{can be reduced by repeating the DEER and NEETR measurements at a second field orientation, for which the candidate tensors predict different transition frequencies.}}

\section{Density Functional Theory Calculations}\label{sec:smDFT}
\red{The DFT methodology is described in Methods, following the same approach as Ref.~\cite{Tang2023}.} As the defects studied are charge neutral, the supercell size of $10.72$ \AA \ is generally sufficient to avoid interactions between the defect and its periodic images. \blue{The hydrogen-related structures (\cref{tab:SIdefects}) are relaxed without symmetry constraints and with the total spin fixed to $S = 1/2$, consistent with our measurements. This constraint does not guarantee that $S = 1/2$ is the lowest-energy spin state of each structure, which we tested explicitly for VH$^0$ (see below). The nitrogen-related structures (\cref{tab:SIdefectsN}) are relaxed within the symmetry of the initial configuration with the total spin unconstrained, so some of these structures may have $S > 1/2$. WAR9, assigned to X2, is recalculated with the settings used for the hydrogen-related structures and evaluated with both PBE and HSE06.}

The hyperfine matrix is then evaluated from the converged electronic structure as $A_{ij}^I=(A_{\rm iso}^I)_{ij} +(A_{\rm aniso}^I)_{ij}$, a summation of the Fermi contact interaction $(A_{\rm iso}^I)_{ij}$ and the magnetic dipolar interaction $(A_{\rm aniso}^I)_{ij}$:
\begin{equation}
\begin{aligned}
\left(A_{\mathrm{iso}}^I\right)_{i j} & =\frac{2}{3} \frac{\mu_0 \gamma_{\mathrm{e}} \gamma_I}{\left\langle S_z\right\rangle} \delta_{i j} \int \delta_T(\mathbf{r}) \rho_{\mathrm{s}}\left(\mathbf{r}+\mathbf{R}_I\right) d \mathbf{r}, \\
\left(A_{\mathrm{aniso}}^I\right)_{i j} &= \frac{\mu_0}{4 \pi} \frac{\gamma_{\mathrm{e}} \gamma_I}{\left\langle S_z\right\rangle} \int \frac{\rho_{\mathrm{s}}\left(\mathbf{r}+\mathbf{R}_I\right)}{r^3} \frac{3 r_i r_j-\delta_{i j} r^2}{r^2} d \mathbf{r},
\end{aligned}
\end{equation}
where $\rho_{\mathrm{s}}$ is the electron spin density, $\mathbf R_I$ is the nucleus coordinate, $\gamma_{\mathrm{e}}$ and $\gamma_I$ are the gyromagnetic ratios for the electron and nuclear spins, $\langle S_z\rangle$ is the average total electron spin in the polar direction, and $\mu_0$ is the vacuum magnetic permeability. The gyromagnetic ratios of hydrogen and nitrogen in our calculation are taken as $\gamma_{\mathrm{n}}(^1\text{H})=42.577$ $\text{MHz/T}$ and $\gamma_{\mathrm{n}}(^{15}\text{N})=-4.316$ $\text{MHz/T}$\blue{, and $\gamma_{\mathrm{n}}(^{14}\text{N})=3.077$~MHz/T for the NV$^-$ benchmark}. 

In order to search the configurational space of the possible defects, we explored different combinations of nitrogen substitutional and interstitial defects with \blue{zero to two} vacancies at positions up to the 8th-nearest neighbor site. \blue{For hydrogen, we \recent{enumerated every symmetry-distinct configuration of a single hydrogen atom with zero to two vacancies near it:
\begin{itemize}\setlength{\itemsep}{0pt}
\item hydrogen on a vacated lattice site with up to two additional vacancies within the third-neighbor shell of that site,
\item hydrogen at a bond-centered, tetrahedral, hexagonal, or antibonding interstitial position with up to two vacancies within 3~\AA, and
\item hydrogen complexes with a $\langle 001\rangle$-split carbon self-interstitial.
\end{itemize}
We} additionally considered hydrogen bonded to the outer carbons of the V--C--V divacancy. \recent{Vacancies farther from the hydrogen were not considered, since the large dipolar coupling of X1 requires electronic spin density close to the proton. The 173} initial configurations relax to \recent{95 distinct structures, which are listed in Supplementary Data~1. Relaxed structures are counted as one if their energies agree within 5~meV and their $^1$H principal values within 0.5~MHz. The} hydrogen-related structures are labeled by their relaxed geometry\recent{, and \cref{tab:SIdefects} lists the results of the lowest-energy structure of each composition together with the structures discussed in the text}. The nitrogen-related structures are labeled by their initial configuration (\cref{tab:SIdefectsN}).}

\blue{We benchmark the accuracy of the hyperfine calculations against \recent{three} defects with known hyperfine tensors. \recent{For NV$^-$, HSE06 gives $^{14}$N hyperfine components of $A_{\parallel}=-2.60$~MHz and $A_{\perp}=-2.94$~MHz and PBE gives $-1.76$ and $-2.10$~MHz, compared with the measured $-2.14$ and $-2.70$~MHz~\cite{Auzinsh2019}. Both functionals thus reproduce the nitrogen hyperfine components to within $\approx 20\%$.} \recent{For a proton next to a single carbon dangling bond, the regime relevant for X1, HSE06 gives $(-16.7, -20.7, +26.7)$~MHz for the N$_2$VH$^0$ (WAR13) center, within 1~MHz of the measured $(-16.1, -20.7, +27.4)$~MHz~\cite{Hartland2014}, whereas PBE gives $(-9.8, -13.6, +29.2)$~MHz.} \recent{HSE06 also reproduces the measured $^1$H tensor of NVH$^-$, in which one unpaired electron is shared by two carbon dangling bonds and the hydrogen tunnels between three equivalent positions~\cite{Glover2003}. Averaged over these positions, HSE06 gives $A_\parallel = -14.7$~MHz and $A_\perp = +7.2$~MHz and PBE gives $-15.5$ and $+5.7$~MHz, compared with the measured $+13.69$ and $-9.05$~MHz, which are known up to an overall sign. HSE06 thus agrees to within 2.1~MHz.} We therefore evaluate the hyperfine tensors of the hydrogen-related structures with HSE06.} We compare our experimental measurements to each defect structure based on the distance to the principal values of the hyperfine matrix in the three-dimensional $(A_1, A_2, A_3)$ space, defined by $d_A = \sqrt{(\langle A_{\perp}^{\mathrm{c}}\rangle - A_{\perp})^2 + (A_\parallel^{\mathrm{c}}-A_{\parallel})^2}$ where $\langle A_{\perp}^{\mathrm{c}}\rangle = (A_1^{\mathrm{c}} + A_2^{\mathrm{c}})/2$ and $A_\parallel^{\mathrm{c}} = A_3^{\mathrm{c}}$. 

\blue{\textit{Sensitivity to the functional and spin state.} All vacancy--hydrogen complexes in \cref{tab:SIdefects} retain several carbon dangling bonds, with spin densities of $\pm(0.2$--$0.45)$  on several carbon atoms. For these structures, PBE and HSE06 converge to different spin configurations, and the $^1$H hyperfine tensor changes by tens of MHz between them. For V$_2$H$^0$ \recent{on its PBE geometry}, for example, PBE gives $(+1.3, +2.3, -4.6)$~MHz and HSE06 $(+9.3, +32.1, -35.2)$~MHz. The calculated tensors of these structures \recent{therefore depend on} the functional and the spin configuration\recent{. The benchmarks above test} the accuracy of HSE06 only for a \recent{single unpaired electron on one or two dangling bonds} (N$_2$VH$^0$\recent{, NVH$^-$}). \recent{For these structures, \cref{tab:SIdefects} lists the tensor of the lowest-energy calculated configuration, which in general is not the state closest to X1. Only the last column of the table comes from the analysis of all $S = 1/2$ states described below.}}

\blue{\recent{\textit{All spin-$1/2$ states of the dangling-bond electrons.} A single DFT calculation of such a structure converges to one broken-symmetry spin configuration~\cite{Noodleman1981}. We therefore evaluate the $^1$H tensor for every $S = 1/2$ state of the dangling-bond electrons with a model of additive hyperfine contributions~\cite{Bencini1990, Mouesca1995}, in which the $n$ radical sites are treated as spins $1/2$. The radical sites are the under-coordinated carbon atoms that carry spin in the HSE06 calculation. The $^1$H tensor of any state is then $\mathbf A = \sum_i K_i \mathbf a_i$, where $\mathbf a_i$ is the tensor for one unpaired electron on site $i$ and $K_i = \langle s_{z,i}\rangle/\langle S_z\rangle$ is the spin projection coefficient of that site. We obtain the $\mathbf a_i$ from a least-squares fit to at least $n+1$ broken-symmetry HSE06 calculations with $S_z = 1/2$ and different spin configurations, for which $K_i = \pm 1$. The residual of the fit tests the additivity of the model. In a second fit, $K_i = \pm 1$ is replaced by the spin moment of site $i$ in each calculation, divided by its mean magnitude over the calculations. This accounts for spin that is not fully localized on the radical sites. We report the closer of the two results and the larger of the two residuals. For a state $\psi$ with $S = 1/2$, the projection coefficients are $K_i = 2\langle\psi|s_{z,i}|\psi\rangle$, and we search all such states of the $n$ spins, including their incoherent mixtures, for the tensor closest to X1. We consider X1 reproduced if a state meets three criteria:
\begin{enumerate}\setlength{\itemsep}{0pt}
\item $d_A \le 5$~MHz, about twice the HSE06 error for the two $^1$H benchmarks,
\item a difference of at most 2~MHz between the two like-sign principal values, compared with $\lesssim 1$~MHz required by the two ZF-DEER transitions of X1 (\cref{sec:smResid}), and
\item a unique axis within $5^\circ$ of the fitted orientation range.
\end{enumerate}
To allow for the error of the additive model, we widen the $d_A$ tolerance by the fit residual. A structure is then counted as not reproducing X1 if no allowed state from either fit meets all three criteria. Since all $S = 1/2$ states are included regardless of their energy, this result does not depend on the spin configuration to which an individual calculation converges. The exceptions are 16 structures for which the calculations do not provide enough independent spin configurations to determine all $\mathbf a_i$. In 15 of them, one or two pairs of radical sites have opposite spins in all broken-symmetry calculations, including calculations started with the two spins parallel. In 12 structures, the two carbons of such a pair are bonded to each other at 1.39--1.44~\AA, shorter than the 1.54~\AA\ bond of diamond and consistent with partial bonding between the two dangling bonds. In the other three, they are second neighbors (2.30--2.54~\AA). Since the pair is antiparallel in every calculation, the fit determines only the difference of its two site tensors, and the tensor of a state with net spin on the pair remains undetermined. We therefore search only the states without net spin on the pair, $K_a + K_b = 0$. For a pair locked in a singlet by a strong antiferromagnetic coupling, as the calculations indicate, the low-energy $S = 1/2$ states satisfy this exactly. The restriction is nevertheless an assumption, since states with net spin on the pair can be closer to X1. V$_2$H$^0$ shows how much the restriction can matter. On its PBE geometry, its pair of radical sites is also antiparallel in all calculations. On its HSE06-relaxed geometry, two calculations have the pair parallel, 0.29 and 0.31~eV above the lowest state, so that all its site tensors are determined. There, the closest allowed state deviates from X1 by 16.9~MHz, compared with 29.8~MHz when the search is restricted to states without net spin on the pair. In the remaining structure, and in one of the 15, the undetermined combination involves more than two sites. There, we search only the states for which the tensor is determined, whose projection coefficients lie in the span of those of the calculated configurations. For all 16 structures, the result therefore applies only to the states searched.}}

\providecommand{\twoline}[1]{\begin{tabular}[t]{@{}l@{}}#1\end{tabular}}%
\begin{table}[ht]
\centering
\blue{\footnotesize\setlength{\tabcolsep}{3pt}
\resizebox{\textwidth}{!}{\begin{tabular}{llccccc}
\toprule
Relaxed structure & Initial configuration(s) & Sym. & $A_1$ $(\theta_1, \phi_1)$ & $A_2$ $(\theta_2, \phi_2)$ & $A_3$ $(\theta_3, \phi_3)$ & \recent{closest $d_A$}\\\midrule
\multicolumn{7}{l}{\textit{Vacancy--hydrogen complexes}}\\
\twoline{VH$^0$, Jahn--Teller\\distorted\recent{$^\dagger$}} & \recent{\twoline{H-sub 0V,\\H-int 1V 1NN, 2NN (+8)}} & C$_{1h}$ & \recent{$+13.52$ (45, $-90$)} & \recent{$+30.06$ (74, 163)} & \recent{$-48.83$ (50, 59)} & \recent{17.6}\\
\quad\twoline{second $S = 1/2$ state\\(\recent{$+51$}~meV)} & \recent{same geometry} & C$_{1h}$ & \recent{$+17.39$ (85, 46)} & \recent{$+36.36$ (22, 149)} & \recent{$-45.05$ (69, $-46$)} & \\
\quad $S = 3/2$ state (\recent{$+0.39$}~eV) & \recent{same geometry} & C$_{1h}$ & \recent{$+0.35$ (45, $-90$)} & \recent{$-0.87$ (65, 27)} & \recent{$+21.73$ (55, 136)} & \\
bond-centered H next to V & H-int 1V 3NN & C$_1$ & $-13.18$ (56, 45) & $+103.12$ (81, $-51$) & $+117.68$ (35, $-154$) & \recent{14.6$^\P$}\\
V$_2$H$^0$\recent{$^\dagger$} & \recent{\twoline{H-sub 1V 1NN,\\H-int 2V V$_2$H (+19)}} & C$_{1h}$ & \recent{$+9.00$ (56, 61)} & \recent{$+34.04$ (68, $-44$)} & \recent{$-37.21$ (42, $-160$)} & \recent{16.9}\\
(V--C--V)H$^0$, H on C0 & \recent{\twoline{H-sub 1V 2NN,\\H-int 2V V-CH-V,\\WAR2 (+14)}} & C$_{1h}$ & $+10.47$ (56, $-39$) & $+17.65$ (34, 146) & $-33.07$ (88, 53) & \recent{27.6}\\
\twoline{(V--C--V)H$^0$, H on\\in-plane outer C} & -- & C$_{1h}$ & $-1.73$ (52, 55) & $-31.86$ (73, $-49$) & $+34.24$ (42, $-158$) & \recent{*}\\
\twoline{(V--C--V)H$^0$, H on\\out-of-plane outer C} & \recent{-- (+5)} & C$_1$ & \recent{$-12.86$ (55, 138)} & \recent{$+25.26$ (83, 43)} & \recent{$-30.19$ (36, $-57$)} & \recent{30.1$^\P$}\\
\recent{\twoline{(V--C--V)H$^0$, H in a\\back bond of C0$^\dagger$}} & \recent{\twoline{H-int BC 2V (both 1NN\\of one bond C)}} & \recent{C$_{1h}$} & \recent{$+23.68$ (60, 36)} & \recent{$-52.64$ (45, $-90$)} & \recent{$-76.00$ (60, 145)} & \recent{0.0}\\
V$\cdots$V(3NN)H$^0$ & \recent{H-sub 1V 3NN (+4)} & C$_{1h}$ & \recent{$+10.73$ (55, 162)} & \recent{$-30.44$ (48, 32)} & \recent{$+34.84$ (62, $-86$)} & \recent{15.8}\\
V$\cdots$V(4NN)H$^0$ & H-sub 1V 4NN & C$_{1h}$ & \recent{$+17.25$ (45, $-90$)} & \recent{$+35.06$ (76, 166)} & \recent{$-58.00$ (48, 64)} & \recent{10.8$^\S$}\\
\recent{V$_3$H$^0$ (1NN, 2NN)$^\dagger$} & \recent{\twoline{H-sub 2V 1NN+1NN,\\1NN+2NN}} & \recent{C$_{1h}$} & \recent{$+10.24$ (90, $-135$)} & \recent{$-29.16$ (90, $-45$)} & \recent{$+29.22$ (0, 119)} & \recent{12.1}\\\midrule
\multicolumn{7}{l}{\recent{\textit{Structures without vacancies}}}\\
\twoline{$\langle 001\rangle$-split I,\\H on a dumbbell C} & \recent{\twoline{H on dumbbell C,\\C--H--C bridge}} & C$_{1h}$ & \recent{$+0.04$ (90, 135)} & \recent{$-3.40$ (50, $-135$)} & \recent{$+16.42$ (40, 45)} & \recent{32.8}\\
\twoline{$\langle 001\rangle$-split I,\\H on a neighboring C} & H at a C--C bond next to I & C$_1$ & \recent{$-0.37$ (83, 20)} & \recent{$-9.86$ (7, $-151$)} & \recent{$+25.92$ (89, 111)} & \recent{24.3}\\
bond-centered H$^0$ & H--C pair on one site & D$_{3d}$ & \recent{$+35.77$ (55, $-135$)} & \recent{$-196.81$ (35, 45)} & \recent{$-196.82$ (90, 135)} & \recent{159}\\
\bottomrule
\end{tabular}}}
\caption{\blue{Hyperfine matrices $\mathbf A$ of the hydrogen-related structures (charge neutral, $S = 1/2$ unless noted), with principal values in MHz. \recent{The table lists the lowest-energy structure of each composition} and the \recent{structures discussed in the text, and all 95 structures are listed in Supplementary Data~1.} Structures are named by their relaxed geometry: V is a vacancy, V$_2$ a nearest-neighbor vacancy pair, V--C--V two vacancies sharing one carbon (C0), and V$\cdots$V($n$NN) a more distant vacancy pair. \recent{The initial configurations that relax to each structure are listed: \textit{H-sub} (\textit{H-int}) denotes hydrogen on a vacated lattice site (at an interstitial site, BC for bond-centered), followed by the number of additional vacancies and their neighbor shells. Shells of two vacancies are joined by $+$, commas separate alternative configurations, and V$_2$H and V-CH-V denote interstitial hydrogen in a divacancy and between two vacancies sharing one carbon.} The two configurations with hydrogen on an outer carbon of (V--C--V)H$^0$ were constructed directly \recent{(marked --)}\recent{, and (+$n$) gives the number of further starting configurations of the systematic search (see text) that relax to the same structure}. The table lists signed principal values and principal directions given by polar angle $\theta$ from the [001] crystallographic direction and azimuthal angle $\phi$ in the (001) plane from [100] towards [010]. Values are HSE06 single points on PBE geometries\recent{, except} for the \recent{structures marked $^\dagger$, which were relaxed with HSE06.} \recent{For the vacancy--hydrogen complexes, the listed values are those of the lowest-energy calculated broken-symmetry spin configuration (see text).} \recent{The last column gives the closest distance $d_A$ to either solution of X1 (MHz), taken over all $S = 1/2$ states of the dangling-bond electrons for the vacancy--hydrogen complexes and for the listed tensor for the structures without vacancies. In the last column, * marks the structure to which this analysis could not be applied, $^\S$ marks a structure for which it is inconclusive, and $^\P$ a structure analyzed for the determined states only (see text).}}}
\label{tab:SIdefects}
\end{table}

\begin{table}[ht]
    \centering
    \begin{tabular}{|c||c||c|c|c|c|c|c|c|c|}
        \hline
           Type&$N_{\rm vacancy}$&defect & $A_{1}$ (MHz)& ($\theta_1$, $\phi_1$) & $A_{2}$ (MHz)& ($\theta_2$, $\phi_2$) & $A_{3}$ (MHz)& ($\theta_3$, $\phi_3$)  &DFT functional\\\hline \hline
 N-substitution& 1& 1NN& 2.95 &  (66, -153) & 2.51 &  (125, 135) & 2.95 &  (45, 90) &GGA-PBE\\\hline
 N-substitution& 1& 2NN& -8.09 &  (90, 135) & -7.7 &  (35, -135) & -9.73 &  (55, 45) &GGA-PBE\\\hline
 N-substitution& 1& 3NN& -13.01 &  (88, 178) & -11.61 &  (45, -90) & -14.41 &  (45, 85)  &GGA-PBE\\\hline
 N-substitution& 1& 4NN& 0.34 &  (135, -90) & 0.01 &  (90, 0) & 0.43 &  (45, -90) &GGA-PBE\\\hline
 N-substitution& 1& 5NN& -18.24 &  (90, 135) & -18.19 &  (55, -135) & -24.59 &  (35, 45)   &GGA-PBE\\\hline
 N-substitution& 1& 6NN& 0.25 &  (45, 90) & 0.1 &  (72, -161) & 0.27 &  (50, -56) &GGA-PBE\\\hline
 N-substitution& 1& 7NN & -0.56 &  (67, -155) & -0.34 &  (54, -47) & -0.61 &  (45, 90) &GGA-PBE\\\hline
 N-substitution& 1& 8NN& -3.51 &  (90, 135) & -3.46 &  (67, -135) & -4.35 &  (23, 45) &GGA-PBE\\\hline
 N-interstitial& 0& \textbf{WAR9}& \blue{11.31} &  \blue{(0, 0)} & \blue{9.35} &  \blue{(90, 45)} & \blue{13.31} &  \blue{(90, 135)} &\blue{GGA-PBE}\\\hline
 & & \textbf{WAR9}& \blue{15.40} &  \blue{(0, 0)} & \blue{13.02} &  \blue{(90, 45)} & \blue{16.85} &  \blue{(90, 135)} &\blue{HSE06}\\\hline
 N-interstitial& 1& 1NN& -112.19 &  (66, -153) & -112.19 &  (55, -45) & -154.91 &  (45, 90)  &GGA-PBE\\\hline
 N-interstitial& 1& 2NN& 422.64 &  (0, 0) & 421.95 &  (90, -135) & 422.83 &  (90, 135) &GGA-PBE\\\hline
 N-interstitial& 1& 3NN& 45.36 &  (8, 45) & 45.13 &  (90, 135) & 75.06 &  (82, -135)  &GGA-PBE\\\hline
  N-interstitial& 1& 4NN& 5.11 &  (114, -27) & -4.45 &  (125, -135) & 5.11 &  (45, -90) &GGA-PBE\\\hline\hline
    \end{tabular}
    \caption{DFT calculations of the hyperfine matrix $\mathbf A$ \red{for possible charge-neutral} \blue{nitrogen-related} \red{defects in diamond}\blue{. Labels describe the initial configuration: nitrogen on (substitution) or off (interstitial) a lattice site, with $N_{\rm vacancy}$ additional vacancies at the listed neighbor shell. These structures were relaxed within the symmetry of the initial configuration, with the total spin unconstrained, except WAR9, which was recalculated without symmetry constraints with $S = 1/2$ and is listed for both PBE and HSE06.} The table lists the \blue{signed principal} values and principal directions \blue{as in~\cref{tab:SIdefects}}. \blue{The WAR9 structure assigned to X2 is highlighted in bold.}}
    \label{tab:SIdefectsN}
\end{table}

\subsection{\blue{Assignment of X2}}
\blue{Since only the relative sign of the experimental hyperfine components is determined (\cref{sec:smResid}), we compare the calculated signed principal values to the experimental tensor for both overall signs. For X2, the closest match is the nitrogen interstitial N$_\text{I}^0$, with calculated principal values of \recent{9.4, 11.3, and 13.3~MHz (PBE) and 13.0, 15.4, and 16.9~MHz (HSE06)}, all of the same sign, consistent with the preferred same-sign solution $(A_\parallel, A_\perp) = (16(1), 6(1))$~MHz. This structure corresponds to the WAR9 defect characterized in Ref.~\cite{Atumi2013}. Our PBE values agree closely with the calculations of Ref.~\cite{Atumi2013}, and our experimental and calculated values have a similar level of consistency to theirs, supporting the assignment. The calculated deviation between the two smallest components of WAR9 ($\approx 2$~MHz) is within the estimated accuracy of the calculation, consistent with the approximately uniaxial hyperfine interaction observed for X2.}

\subsection{\blue{Comparison of X1 with calculated structures}}
\blue{None of the \recent{calculated tensors of the 95 distinct} hydrogen-related structures (\cref{tab:SIdefects} \recent{and Supplementary Data~1}) reproduces the signed hyperfine tensor of X1 for either of the two opposite-sign solutions. \recent{By composition, 23 of these structures have three vacant lattice sites, 64 have two, 4 have one, and 2 have none. The other 2 are hydrogen complexes with a carbon self-interstitial. The lowest-energy structures of the compositions with one, two, and three vacant sites are VH$^0$, V$_2$H$^0$, and V$_3$H$^0$ (1NN, 2NN), respectively. Every single-configuration tensor in \cref{tab:SIdefects}} either fails to reproduce the two observed \recent{ZF-DEER} transitions of X1 at 7.0 and 31.8~MHz or predicts additional transitions within the scanned range of 1--35~MHz that are not observed for X1 \recent{(\cref{sec:smResid}).}}

\blue{\recent{Five structures have a single unpaired electron. In three of them, the two self-interstitial complexes with hydrogen on a carbon and one structure with three vacant lattice sites, the proton is next to a carbon dangling bond, as in the N$_2$VH$^0$ and NVH$^-$ benchmarks, for which the HSE06 tensor is accurate to $\approx 2$~MHz. These three deviate from X1 by $d_A \geq 24$~MHz. Bond-centered and tetrahedral H$^0$ lie outside this regime but deviate from X1 by more than 100~MHz.}}

\blue{\recent{The analysis applies to 65 structures with several dangling bonds, of which 49 are analyzed for all $S = 1/2$ states and 16 for the determined states only. For 47 of the 49, no state reproduces X1 within the tolerances above, and the other two are inconclusive (see below). For most of them, the closest allowed state deviates from X1 by $d_A > 10$~MHz, for example by 17.6~MHz for the} Jahn--Teller-distorted VH$^0$\recent{, 16.9~MHz for V$_2$H$^0$, 12.1~MHz for V$_3$H$^0$ (1NN, 2NN), and 27.6~MHz for (V--C--V)H$^0$ with H on C0 (\cref{tab:SIdefects}). These values, from the last column of \cref{tab:SIdefects}, ignore the rhombicity and axis criteria and are therefore lower bounds on the deviation from X1. For the other 16, no determined state reproduces X1, and the closest deviates by $d_A \geq 11.0$~MHz. These include the only two structures within 1~eV of the lowest-energy structure of their composition, 0.10 and 0.43~eV above V$_3$H$^0$ (1NN, 2NN). For these two, the conclusion rests on the assumption that the antiparallel pair carries no net spin.}}

\blue{\recent{The residual of the additive model is below 1~MHz for 34 of the 65 tested structures, 1--3~MHz for 12, and 3.3--9.7~MHz for the other 19. The closest case is (V--C--V)H$^0$ with H in a back bond of C0 (\cref{tab:SIdefects}), 5.6~eV above the lowest-energy structure of its composition. On its HSE06-relaxed geometry, one allowed state comes within $d_A = 1.9$~MHz of X1, with 0.5~MHz between its like-sign principal values, but its unique axis lies $7^\circ$ outside the fitted orientation range. Since the fitted orientation depends on the alignment of the magnetic field along [111], we also test the effect of relaxing the orientation criterion. This structure would reproduce X1 with an axis tolerance of $6.4^\circ$ or more, through a state at the limits of the other two criteria ($d_A = 5.0$~MHz and 2.0~MHz between the like-sign principal values). For all other structures, dropping the orientation criterion entirely does not change the verdict. With all three tolerances doubled, no other tested structure would reproduce X1, and neither would any listed tensor of the 25 structures for which this analysis could not be applied or was limited to a single spin configuration (see below).}}

\blue{\recent{The relaxation with HSE06 matters for V$_3$H$^0$ (1NN, 2NN), in which a dangling bond points at the proton from 1.89~\AA\ in the PBE geometry. On this geometry, the closest allowed state comes within $\approx 5$~MHz of the magnitude} of solution (\recent{ii), with a non-uniaxial tensor and a principal axis outside the fitted orientation range. The HSE06 relaxation moves the hydrogen 0.08~\AA\ away from this dangling bond and increases the deviation to 12.1~MHz. The values above for VH$^0$, V$_2$H$^0$ and the closest structure are also on HSE06-relaxed geometries, whereas all other structures, like the two benchmarks, are evaluated on PBE geometries. Compared with the PBE geometry, the relaxation changes the closest $d_A$ by $-5$~MHz for the closest structure, $+7$~MHz for V$_3$H$^0$ (1NN, 2NN), and less than 1~MHz for VH$^0$ and V$_2$H$^0$ (for V$_2$H$^0$ with the same restriction of the states on both geometries).}}

\blue{\recent{The analysis could not be applied to 19 structures, all at least 2.8~eV above the lowest-energy structure of their composition. For (V--C--V)H$^0$ with H on the in-plane outer carbon, the broken-symmetry calculations converge to a single spin configuration, and} its tensor is strongly non-uniaxial \recent{(\cref{tab:SIdefects}). For three structures, the additive model fails, with residuals of 10--46~MHz. For six, most calculations converge to spin configurations that do not correspond to $S_z = 1/2$ on the radical sites identified from the PBE calculation, which leaves too few usable configurations. For nine structures, 3.0--9.1~eV above the lowest-energy structure of their composition, the number of radical sites is even, and their single-configuration tensors deviate from X1 by $d_A \geq 6.5$~MHz. For six further structures, 3.4--6.3~eV above the lowest-energy structure of their composition, only the HSE06 tensor of a single spin configuration was calculated, and it deviates from X1 by $d_A \geq 12.8$~MHz. For V$\cdots$V(4NN})H$^0$ \recent{and for a structure 3.55~eV above the lowest-energy structure of its composition, the analysis is inconclusive. An allowed state meets the criteria within the fit residual (7.1 and 9.0~MHz), which arises from a partial pairing of radical sites in some of the calculations. For VH$^0$, calculations started from different} spin configurations, including single-radical and high-spin ($S = 3/2$) starting points\recent{, converge} to one of two $S = 1/2$ states separated by \recent{51}~meV \recent{on the HSE06-relaxed geometry, and the} $S = 3/2$ state lies \recent{0.39}~eV higher \recent{(\cref{tab:SIdefects}). The second $S = 1/2$ state has a spin density of lower symmetry than the geometry.}}

\blue{\recent{These results rely on the additive model, the HSE06 couplings, and, except for the structures marked $^\dagger$ in \cref{tab:SIdefects}, the PBE geometries. Given the number of structures, we did not test the convergence of the calculated hyperfine tensors with respect to supercell size or k-point sampling ($\Gamma$-only), and we did not include zero-point or thermal motion of the proton, which could average the tensor between nearby configurations at room temperature. In addition, our search does not cover structures with more than one hydrogen atom, more vacancies near the hydrogen, or other impurities,} so the absence of a match does not exclude \recent{a vacancy--hydrogen complex outside this set}. }

\blue{We also calculated hydrogen complexes with a carbon self-interstitial, which is formed during ion implantation. Hydrogen bonded to one carbon of the $\langle 001\rangle$-split self-interstitial leaves a single radical carbon, but gives small hyperfine couplings (\recent{$T \lesssim 10$}~MHz) in both configurations considered (\cref{tab:SIdefects})\recent{. A C--H--C bridge between two carbons of the interstitial relaxes to the same structure as hydrogen on a dumbbell carbon. A} hydrogen--carbon pair sharing one lattice site relaxes to bond-centered H$^0$, with a $\langle 111\rangle$-axial tensor of \recent{$(+35.8, -196.8, -196.8)$}~MHz. None of these structures reproduces X1.}

\blue{The WAR2 center detected with EPR was proposed to be a V-CH-V$^0$ complex with the hydrogen on the $\langle 100\rangle$ bisector of the two vacancies~\cite{Cann2009}. \recent{In our calculations, this geometry is unstable and relaxes to (V--C--V)H$^0$ with the C--H bond tilted by $25^\circ$ from $\langle 100\rangle$ within the mirror plane. As in the calculations of Ref.~\cite{Atumi2020}, its hyperfine tensor does not reproduce the WAR2 EPR values (\cref{tab:SIdefects}).} The hyperfine components reported for WAR2, $(\mp20.1, \pm30.9, \mp27.7)$~MHz along [001], [110], and [1$\bar{1}$0]~\cite{Cann2009}, share the opposite-sign pattern of X1\recent{. With $T\approx 18$~MHz and a small isotropic contribution, $a_{\text{iso}}\approx\mp5.6$~MHz, similar to solution (i) of X1 ($a_{\text{iso}}\approx\mp4.3$~MHz), WAR2 is the closest known hydrogen-related center.} However, the WAR2 tensor is strongly non-uniaxial, with a difference of 7.6~MHz between its two same-sign components, and would give four ZF-DEER transitions at 1.6, 5.4, 25.5, and 29.3~MHz (\cref{tab:SIliterature}), inconsistent with the two transitions observed for X1 at 7.0 and 31.8~MHz. Independently of the ZF-DEER spectrum, the WAR2 tensor is also inconsistent with the DEER and NEETR frequencies of X1\recent{. For} a field along [111], its two magnetically distinct orientations give hyperfine splittings of 25.4 and 27.9~MHz and nuclear splittings of 2.74 and 1.85~MHz, compared with the measured 26.4 and 2.217~MHz\recent{. The discrepancy remains} for field misalignments of up to $5^\circ$. In addition, the unique [110] axis of WAR2 lies at either $35^\circ$ or $90^\circ$ from the [111] field direction, compared with $\vartheta\approx 78^\circ$ or $24^\circ$ for X1. We therefore conclude that X1 is distinct from WAR2.}

\subsection{\blue{Comparison of X1 with known hydrogen-related centers}}
\blue{We further compare X1 to hydrogen-related paramagnetic centers in diamond characterized with EPR, and to hydrogen-related structures calculated in previous DFT studies~(\cref{tab:SIliterature}). For each center, we list the signed principal values as reported, the isotropic and dipolar contributions $a_{\text{iso}}$ and $T$, and the ZF-DEER transition frequencies predicted from the formulas of \cref{sec:smZFDEER}. X1 shows only two ZF-DEER transitions, at 7.0 and 31.8~MHz, and none of the listed centers reproduces them. All previously reported hydrogen-related centers with signed tensors have dipolar contributions $T \lesssim 18$~MHz, whereas X1 requires $T\approx 21$~MHz. \recent{The centers with the most similar tensor shape, with opposite-sign principal values and a small isotropic contribution as in solution (i) of X1, have a C--H bond next to a single remaining carbon dangling bond: N$_2$VH$^0$, VH$_3^0$, NVH$_2^0$, and OVH$_2^+$~\cite{Hartland2014, Peaker2018, Thiering2016}.} However, their dipolar contributions are 17--34\% smaller than for X1, their tensors are more strongly non-uniaxial, and several contain additional nuclear spins (nitrogen or additional hydrogen) that would produce resolvable hyperfine couplings. Isolated bond-centered hydrogen, which in diamond has been characterized through its muonium analog~\cite{Holzschuh1982}, is also excluded: scaled to the proton gyromagnetic ratio, the muonium tensor is axial along $\langle 111\rangle$, with components several times larger than those of X1. Our calculated bond-centered H$^0$\recent{, $(+35.8, -196.8, -196.8)$~MHz with HSE06 (\cref{tab:SIdefects}), is likewise axial along $\langle 111\rangle$ with much larger components. A quantitative comparison with the scaled muonium tensor is limited by the large zero-point motion of the muon}.}
\begin{table}[ht]
\centering
\blue{\footnotesize
\begin{tabular}{lllccll}
\toprule
Center & Method & Principal values (MHz) & $a_{\text{iso}}$ & $T$ & ZF-DEER lines (MHz) & Notes\\\midrule
X1, solution (i) & this work & $-25.8,\,-25.8,\,+38.7$ & $-4.3$ & 21.5 & 6.5, 32.3 & \\
X1, solution (ii) & this work & $-38.4,\,-38.4,\,+23.4$ & $-17.8$ & 20.6 & 7.5, 30.9 & \\\midrule
VH$^0$, JT (this work) & DFT (HSE06) & \recent{$+13.5,\,+30.1,\,-48.8$} & \recent{$-1.8$} & \recent{$-23.5$} & \recent{9.4, 17.7, 31.2, 39.4} & several dangling bonds\\
V$_2$H$^0$ (this work) & DFT (HSE06) & \recent{$+9.0,\,+34.0,\,-37.2$} & \recent{$+1.9$} & \recent{$-19.6$} & \recent{1.6, 14.1, 23.1, 35.6} & several dangling bonds\\
(V--C--V)H$^0$ (this work) & DFT (HSE06) & $+10.5,\,+17.7,\,-33.1$ & $-1.7$ & $-15.7$ & 7.7, 11.3, 21.8, 25.4 & several dangling bonds\\
WAR2~\cite{Cann2009} & EPR & $\mp20.1,\,\mp27.7,\,\pm30.9$ & $\mp5.6$ & $\pm18.3$ & 1.6, 5.4, 25.5, 29.3 & non-uniaxial\\
H1~\cite{Zhou1996} & EPR & $-6,\,-6,\,+27$ & $+5.0$ & 11.0 & 10.5, 16.5 & \\
NVH$^-$~\cite{Glover2003} & EPR & $-9.05,\,-9.05,\,+13.69$ & $-1.5$ & 7.6 & 2.3, 11.4 & N hyperfine\\
N$_2$VH$^0$ (WAR13)~\cite{Hartland2014} & EPR & $-16.07,\,-20.68,\,+27.41$ & $-3.1$ & 15.3 & 3.4, 5.7, 21.7, 24.0 & two N\\
VH$_3^0$~\cite{Peaker2018} & DFT (PBE) & $-19,\,-16,\,+27$ & $-2.7$ & 14.8 & 4.0, 5.5, 21.5, 23.0 & three H\\
NVH$_2^0$~\cite{Peaker2018} & DFT (PBE) & $-17,\,-14,\,+27$ & $-1.3$ & 14.2 & 5.0, 6.5, 20.5, 22.0 & N, two H\\
OVH$_2^+$~\cite{Thiering2016} & DFT (HSE06) & $-25.8,\,-23.5,\,+26.3$ & $-7.7$ & 17.0 & 0.2, 1.4, 24.9, 26.1 & two H\\
Mu$_\text{BC}$~\cite{Holzschuh1982} & $\mu$SR, scaled to $^1$H & $|123.3|,\,|123.3|,\,|52.8|$ & -- & -- & 35.2, 88.0 & \\
Other centers~\cite{Iakoubovskii2002, Glover2004, Edmonds2008, DHaenensJohansson2010, DHaenensJohansson2011} & EPR & all $|A|\lesssim 8$ & & & & or $S=1$\\
\bottomrule
\end{tabular}}
\caption{\blue{Comparison of the X1 hyperfine tensor with hydrogen-related centers in diamond. Principal values are listed as reported (signed where available), with $a_{\text{iso}}$ the isotropic and $T$ the dipolar contribution. For X1, the best-fit tensors of \cref{tab:signfit} are listed, with the overall sign chosen such that $A_\parallel > 0$. The ZF-DEER lines predicted for each X1 solution are listed along with the corresponding lines for each center, with four lines for non-uniaxial tensors (\cref{sec:smZFDEER}). The measured X1 lines are at 7.0 and 31.8~MHz. The signs of $a_{\text{iso}}$ and $T$ reverse with the overall sign of a tensor, which is undetermined for X1.}}
\label{tab:SIliterature}
\end{table}

\blue{Based on this comparison, X1 is a hydrogen-related defect distinct from all previously characterized hydrogen centers in diamond. The large dipolar contribution, $T\approx 21$~MHz for both solutions, suggests a proton close to a localized dangling-bond orbital. The fitted orientation of the principal axis (\cref{sec:smResid}) further excludes trigonal centers with a $\langle 111\rangle$ axis, including centers with motionally averaged trigonal symmetry, as well as structures with $\langle 110\rangle$- or $\langle 100\rangle$-oriented principal axes. Candidate structures for future calculations are therefore low-symmetry (C$_{1h}$ or C$_1$) hydrogen-related structures. These include partially hydrogenated vacancy clusters (V$_n$H$_m$) leaving a single radical carbon adjacent to one C--H bond, provided the additional protons couple only weakly to the electronic spin, as well as other low-symmetry vacancy--hydrogen complexes beyond those calculated here.}

\section{Ionization analysis of defect charge stability}\label{sec:smIonization}

\red{In order to assess the charge stability of the charge-neutral WAR9 \blue{defect assigned to X2}, we implement an ionization analysis based on DFT calculations~\cite{squires2026guidelines}. The stability of a defect charge state is determined by the defect formation energy of the charge state relative to other possible charge states. The defect formation energy with a charge state $q$, $\Delta E_{D, q}$, is evaluated by the DFT calculations as~\cite{gorai2025beginner}:
\begin{equation}
    \Delta E_{D, q}=E_{D, q}-E_0+\sum_i n_i \mu_i+E_{\mathrm{corr}}+q (E_{\mathrm{VBM}} + E_{\mathrm{F}}),
\end{equation}
where $E_{D,q}$ is the energy of the diamond supercell with a defect, $E_0$ is the energy of the pristine diamond supercell (without a defect), $\mu_i$ is the chemical potential, and $n_i$ is the difference in the number of atoms between the pristine and defect supercells of each chemical species $i$. $E_{\rm corr}$ is a correction term to account for the unphysical long-range Coulomb interaction of charged defects with their periodic images in the supercell simulations, and $E_{\rm F}$ is the electron Fermi level measured from the valence band maximum $E_{\rm VBM}$. }

\red{To determine the range of $E_{\rm F}$ where the proposed defect charge states are stable, we evaluate the defect formation energies of different charge states as a function of $E_{\rm F}$ for \blue{the WAR9 defect} together with the NV$^-$ defect as a reference. The supercell energies for the system with and without defects, $E_{D,q}$ and $E_0$, are evaluated by the DFT calculations with the computational settings described in the Methods section included in the main text. We evaluate the chemical potential of carbon, $\mu_C$, as the atomic energy of the pristine diamond supercell ($E_0$ divided by the number of atoms), and set the chemical potential of nitrogen to the isolated-atom reference energy. Note that  setting the chemical potential only influences chemical stability and does not influence our conclusion of charge stability of the defects. The correction energy $E_{\rm corr}$ is evaluated by the Kumagai-Oba (eFNV) charge correction scheme~\cite{kumagai2014electrostatics} implemented by the doped Python toolkit~\cite{kavanagh2024doped}, with an isotropic dielectric constant of 5.7.}

\begin{figure}[ht]
	\centering
	\includegraphics[width=4in]{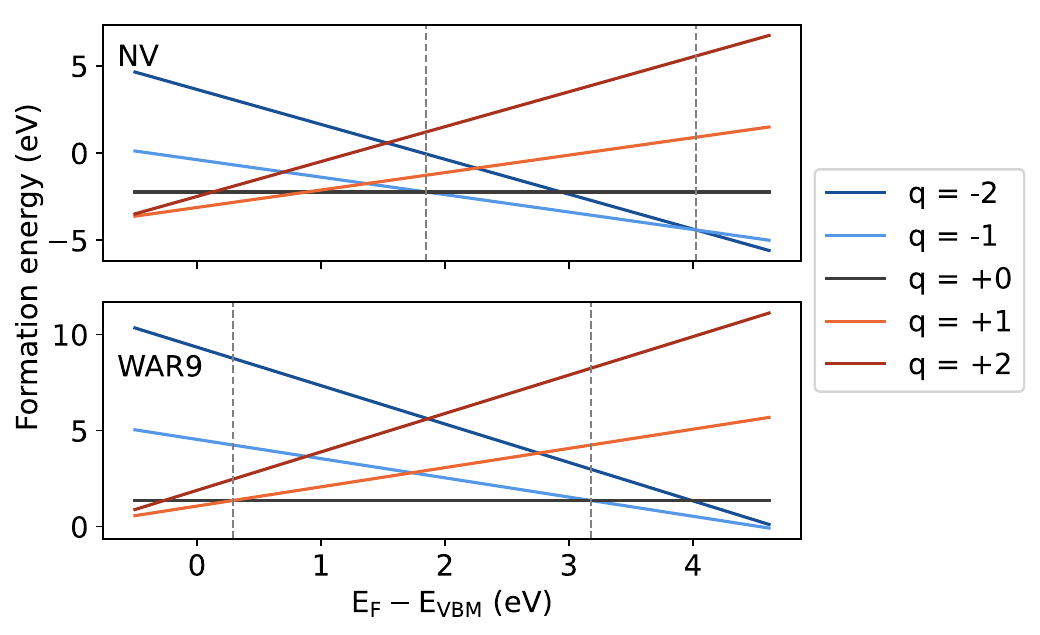}
	\caption{\red{Defect formation energies of the NV$^-$ and WAR9 defects as a function of the electron Fermi energy $E_{\rm F}$. The charge states q = -2, -1, 0, 1, 2 are calculated for each defect. The stability ranges of $E_{\rm F}$ are denoted by the vertical dashed lines for our identified defects (the charge neutral state for \blue{WAR9} and the negatively charged state for NV as a reference). The horizontal axis of electron Fermi energy takes the valence band maximum $E_{\rm VBM}$ as the zero point. The calculated (PBE) band gap is 4.1~eV, smaller than the experimental value of 5.5~eV.}}
	\label{fig:ionization}
\end{figure}

\red{The calculated defect formation energies are shown in~\cref{fig:ionization}. At each $E_{\rm F}$, the thermodynamically stable charge state is the one with the lowest formation energy (here we neglect entropy effects following the common assumptions in the literature~\cite{squires2026guidelines}, so the formation energy approximates the formation free energy). \blue{The ranges of $E_{\rm F}$ in which the q=-1 state of NV and the q=0 state of WAR9 reach the lowest formation energy compared to other charge states are found between the vertical dashed lines. Both defect charge states exhibit wide ranges of $E_{\rm F}$ within the diamond bandgap for charge stability. In addition, there is a common interval from approximately 1.8~eV to 3.2~eV above the valence band maximum where both charge states are simultaneously stable. Therefore, the calculation supports the conclusion that the neutral WAR9 defect can stably coexist with the negatively charged NV center in diamond.}}

\section{Fidelity of X1 nuclear spin polarization}\label{sec:smFidelity}

\begin{figure}[ht]
	\centering
	\includegraphics[width=6.8in]{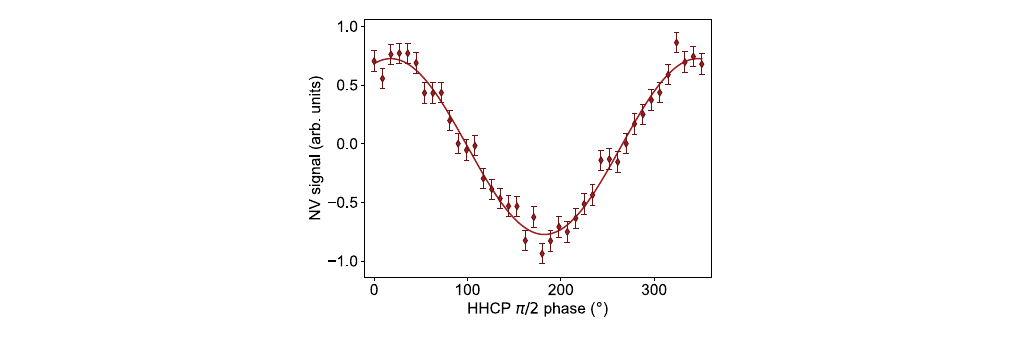}
	\caption{Calibrating the polarization transfer efficiency with Hartmann-Hahn Cross Polarization (HHCP) between the NV and X electronic spins to estimate nuclear polarization fidelity using the sequence in~\cref{fig:control}b. Here we apply two consecutive HHCP steps with simultaneous driving of both X hyperfine transitions to perform initialization and readout of the X electronic spin, while sweeping the phase of the first $\pi/2$-pulse during the readout step. The signal is fit to $y(\phi_{\text{HHCP}}) = a_0\cos(2\pi\phi_{\text{HHCP}}/\phi_0 + \theta_0) + b_0$ where $a_0 = \red{0.75(2)}$, corresponding to an HHCP fidelity of $F_1 = 0.87(1)$.}
	\label{fig:supp_SPAM}
\end{figure}

The protocol for nuclear spin initialization for the X1 defect~(\cref{fig:control}b) consists of two sequential ($i$)SWAP gates. The first $i$SWAP is between the NV and X electronic spins via Hartmann-Hahn Cross Polarization (HHCP), and the second SWAP is between the X electron and X nuclear spins. The electron-nuclear SWAP gate is implemented by applying two conditional electron and nuclear $\pi$-pulses (CNOTs). The amount of nuclear polarization is limited by the NV polarization efficiency ($\eta$), the NV-X HHCP fidelity ($F_1$), and electron-nuclear SWAP fidelity ($F_2$). The NV polarization efficiency is estimated to be approximately 80\% from  previous work~\cite{Robledo2011}, setting an upper bound to the measured hydrogen nuclear polarization. The HHCP fidelity is limited by control error on the X electronic spin from off-resonant driving, as well as from leakage to other near resonant spins~\cite{Ungar2024}. We calibrate the HHCP fidelity by measuring the NV contrast after round-trip polarization transfer between the NV and X electronic spins~(\cref{fig:supp_SPAM}). We apply two sequential HHCP $i$SWAP gates and sweep the phase of the first $\pi$/2-pulse in the readout-$i$SWAP. The NV contrast is estimated from the amplitude of the cosine fit, with $a_0 = \red{0.75(2)}$, leading to an HHCP fidelity of $F_1 = \sqrt{a_0} = 0.87(1)$. 

\begin{figure}[ht]
	\centering
	\includegraphics[width=6.8in]{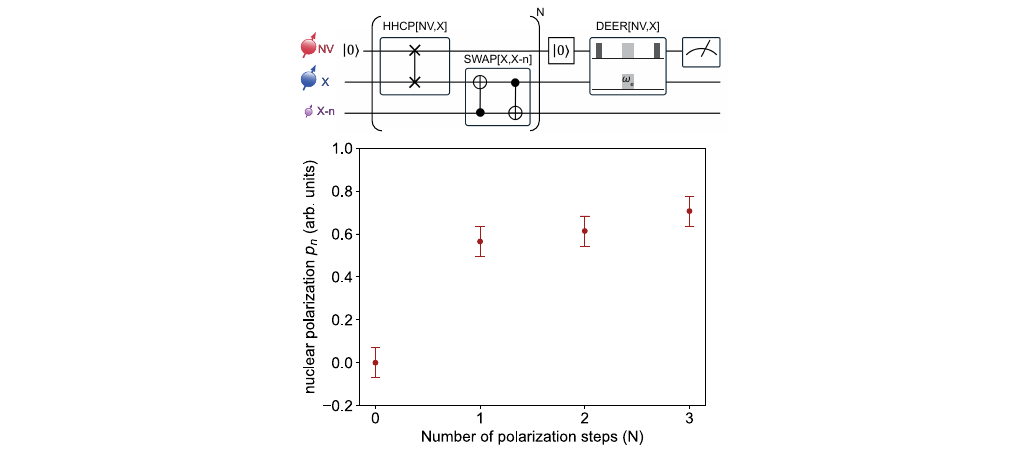}
	\caption{Polarization of the hydrogen nuclear spin after repeating the initialization sequence up to $\text{N}=3$ times. The polarization $p_{\mathrm{n}}$ is estimated from the integrated area of the $\ket{\downarrow_{\mathrm{n}}}$ hyperfine resonance in the DEER measurement (not shown), and scaled to the $\text{N}=1$ value calculated from~\cref{fig:control}b. The nuclear polarization after 3 repetitions is $p_{\mathrm{n}} = \red{0.71(7)}$, consistent with the maximum expected polarization when accounting for imperfections in NV optical polarization and NV-X HHCP. While all three polarization values agree within error, the potential small increase from $\text{N}=1$ to $\text{N}=3$ can be used to estimate the electron-nuclear SWAP fidelity.}
	\label{fig:supp_pol}
\end{figure}

The total nuclear polarization can be estimated from the DEER measurement data in \cref{fig:control}b by calculating the integrated area of each resonance dip. By fitting a single-Lorentzian curve to each resonance (see Methods), we find that the amplitude and HWHMs of each peak after nuclear initialization are $a_{\downarrow}, a_{\uparrow} = \red{1.50(7),\; 0.50(4)}$ and $\gamma_{\downarrow}, \gamma_{\uparrow} = \red{0.67(6), \; 0.56(9)}$ MHz. For the reference without the nuclear initialization step, the amplitude and HWHM are identical across both resonances, with $a_0 = 1.04$ and $\gamma_0 = 0.38$ MHz, indicating zero polarization. The total nuclear polarization is then estimated as the relative difference in the areas under each resonance peak:
\begin{align*}
    p_{\mathrm{n}} = \left(\frac{a_{\downarrow}\gamma_{\downarrow} - a_{\uparrow}\gamma_{\uparrow}}{a_{\downarrow}\gamma_{\downarrow} + a_{\uparrow}\gamma_{\uparrow}}\right) = \red{0.57(7)}.
\end{align*}
This estimate is close to the maximum nuclear polarization expected after accounting for the imperfect NV polarization and NV-X HHCP, which is equal to $\eta \times F_1 \approx 0.7$. Although the estimate for $p_{\mathrm{n}}$ agrees within \red{two standard deviations} with the calculated maximum, any remaining loss can be attributed to the imperfect electron-nuclear SWAP gate. We verify this by repeating the initialization sequence up to $\text{N}=3$ cycles and find the nuclear polarization slightly increases, although the measured values all still agree within error~(\cref{fig:supp_pol}). We observe a maximum nuclear polarization of $p_{\mathrm{n}} = \red{0.71(7)}$ after $\text{N}=3$ repetitions. This suggests the hydrogen spin reaches the calibrated upper-bound in polarization, leading to an estimate in the electron-nuclear SWAP fidelity of $F_2 \approx \red{0.8}$. While the NV optical polarization step is intrinsically limited by the photo-physics of the defect, control error associated with the X electronic and nuclear spins can be reduced with more precise calibration to achieve maximum polarization transfer in a single repetition. 

\section{Coherence of X2 nitrogen nuclear spin}\label{sec:smT2n}
\begin{figure}[ht]
	\centering
	\includegraphics[width=6.8in]{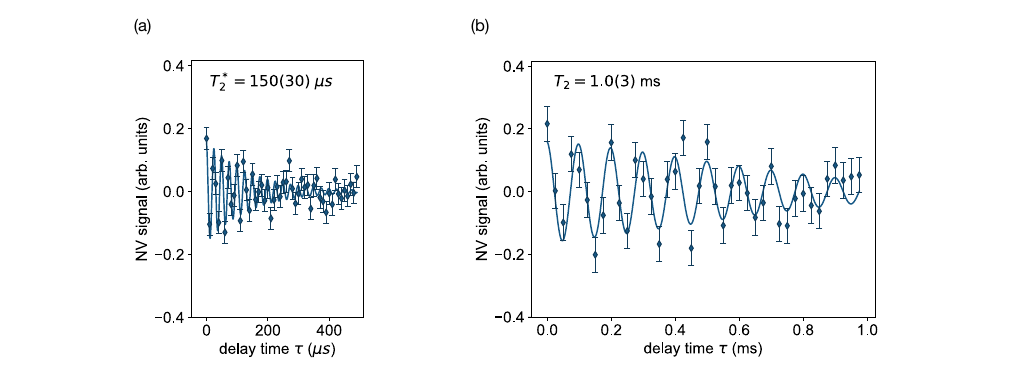}
	\caption{Coherence measurements of the X2 nitrogen nuclear spin qubit using the NEETR pulse sequence (see control protocol in~\cref{fig:NVXENDOR}a). The phase-modulated Ramsey and spin-echo evolution is applied during the conditional $R(\theta)$ block (see insets of~\cref{fig:control}c,d). (a) Measuring nuclear spin dephasing time using phase-modulated Ramsey at $f_{\text{mod}}= 20 \, \text{kHz}$, resulting in $T_2^{*} = 150(30) \, \mu\text{s}$. (b) Measuring nuclear spin coherence time using phase-modulated spin-echo at $f_{\text{mod}} = 5 \, \text{kHz}$, resulting in $T_2 = 1.0(3) \, \text{ms}$. We fit both signals to an exponentially decaying cosine function with decay constant $T = \left(\frac{1}{T_2^{(*)}} + \frac{3}{2T_1^{\mathrm{e}}}\right)^{-1}$ to estimate the intrinsic nuclear coherence times. We find that a \red{Gaussian decay} of the form $y(t) = a\cos(\omega t +\phi)e^{ -(t/T)^{2} }$ yields an improved fit for the spin-echo signal, and use this fit to determine the reported $T_2$. This improved fit is not applied to the X1 spin-echo measurement in \cref{fig:control}d.}
	\label{fig:supp_X2T2}
\end{figure}

We perform the nuclear coherence measurements from \cref{fig:control}c,d on the X2 nitrogen spin and find comparable results to the X1 hydrogen spin for both nuclear Ramsey and spin-echo protocols. The measurements are displayed in \cref{fig:supp_X2T2}, which show $T_2^{*} = 150(30) \, \mu\text{s}$ and $T_2 = 1.0(3)\, \text{ms}$, where we have accounted for the X2 electronic spin relaxation in our estimation of the decay constants (\cref{sec:smT1e}). Together with the hydrogen nuclear spin of X1, the long-lived nuclear coherence times of both defects highlight how our approach can be scaled to construct larger nuclear-spin registers surrounding a central NV spin, without being limited by the direct NV-nuclear coupling.

\section{Electronic spin relaxation effects on nuclear spin coherence measurements}\label{sec:smT1e}
To estimate the intrinsic decoherence times of the nuclear spin due to the bath environment ($T_2^{*}, T_2$) we need to decouple the effects from the strongly coupled electronic spin. Measurement of the nuclear spin coherence with the NEETR sequence~(\cref{fig:control}c,d) relies on initialization and readout of the X electronic spin population via polarization transfer with the NV. As a result, the total measurement contrast decays with time constant $T_1^{\mathrm{e}}$, which is the electronic spin relaxation time. This contribution to the decay of the measured signal can be classified as state preparation and measurement (SPAM) error. Additionally, during free evolution, the nuclear spin will experience a random phase shift from the electronic spin via the strong hyperfine coupling, which serves as a source of decoherence. By modeling the electronic spin fluctuation as random telegraph noise (RTN), the nuclear spin dephasing time is limited by the electronic spin relaxation time in the strong coupling limit, with a decoherence rate of $1/(2T_1^{\mathrm{e}})$~\cite{Chen2018}. The total decay rate of the coherence signal, with decay constant $T$, is equal to the sum of all three individual decoherence rates:
\begin{align*}
    \frac{1}{T} &= \Gamma_\text{Bath} + \Gamma_\text{SPAM} + \Gamma_{\text{RTN}} \\
    &= \frac{1}{T_2^{*}} + \frac{1}{T_1^{\mathrm{e}}} + \frac{1}{2T_1^{\mathrm{e}}}
\end{align*}

\begin{figure}[ht]
	\centering
	\includegraphics[width=6.8in]{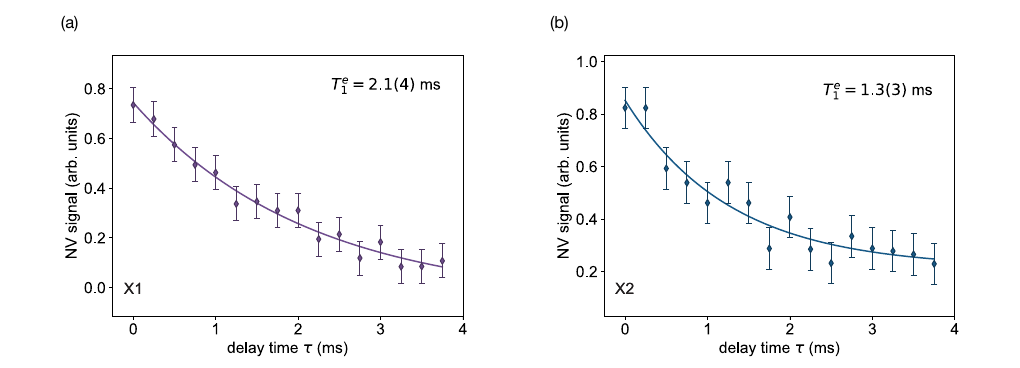}
	\caption{Measuring the electronic spin relaxation time $T_1^{\mathrm{e}}$ for (a) X1 and (b) X2 to account for SPAM and dephasing error on the nuclear spin coherence measurements~(\cref{fig:control}c,d). We apply the NEETR sequence initialization and readout steps separated by a delay time $\tau$ while the RF drive is turned off (sequence in~\cref{fig:NVXENDOR}a). Data is fit to $y(t) = a_0e^{-t/T} + b_0$.}
	\label{fig:supp_T1e}
\end{figure}

To extract the intrinsic nuclear spin coherence times from the measured signal, we fit the data to an exponentially decaying cosine function with time constant $T = (\frac{1}{T_2^{*}} + \frac{3}{2T_1^{\mathrm{e}}})^{-1}$, where $T_{2}^{*}$ is a free parameter and $T_1^{\mathrm{e}}$ is fixed. The values used for $T_1^{\mathrm{e}}$ for the X1 and X2 defects are extracted from the measurements displayed in~\cref{fig:supp_T1e}. Since the hyperfine coupling strength for both defects is much larger than the $T_1^{\mathrm{e}}$ flip rate, both $T_2^{*}$ from the Ramsey measurement and $T_{2}$ from the spin-echo measurement can be estimated using this fitting approach. The estimated X1 hydrogen spin coherence times are reported in the main text~(\cref{fig:control}c,d), and the X2 nitrogen spin coherence times are reported above (\cref{sec:smT2n}).

\putbib[ENDOR_fixed]

\end{bibunit}

\end{document}